\documentclass[fleqn,usenatbib]{mnras}

\usepackage{newtxtext,newtxmath}

\usepackage[T1]{fontenc}

\DeclareRobustCommand{\VAN}[3]{#2}
\let\VANthebibliography\thebibliography
\def\thebibliography{\DeclareRobustCommand{\VAN}[3]{##3}\VANthebibliography}

\usepackage{graphicx}	    
\usepackage{amsmath}	    
\usepackage{gensymb}	    
\usepackage{anyfontsize}    
\usepackage{xcolor}         

\usepackage[normalem]{ulem}

\newcommand{\thisstar}{HD~28185}
\newcommand{\thisstarb}{HD~28185~b}
\newcommand{\thisstarc}{HD~28185~c}
\newcommand\ms{\ensuremath{\text{m}\,\text{s}^{-1}}}
\newcommand\msyr{\ensuremath{\text{m}\,\text{s}^{-1}\,\text{yr}^{-1}}}
\newcommand\masyr{\ensuremath{\text{mas}\,\text{yr}^{-1}}}

\definecolor{my_color}{HTML}{CF0000}                            
\newcommand\bmaroon{}

\defcitealias{GaiaEDR3}{Gaia EDR3}
\defcitealias{Feng2022}{F22}
\defcitealias{WDS}{WDS}

\title[HD 28185 Revisited]{HD 28185 Revisited: An Outer Planet, Instead of a Brown Dwarf,\\ On a Saturn-like Orbit}

\author[Venner et al.]{Alexander Venner,$^{1}$\thanks{E-mail: alexandervenner@gmail.com}
Qier An,$^{2,3}$
Chelsea X. Huang,$^{1}$
Timothy D. Brandt,$^{2,4}$
Robert A. Wittenmyer,$^{1}$
\newauthor{Andrew Vanderburg$^{5}$}
\\
$^{1}$Centre for Astrophysics, University of Southern Queensland, Toowoomba, QLD 4350, Australia \\
$^{2}$Department of Physics, University of California Santa Barbara, Santa Barbara, CA 93106, USA \\
$^{3}$Department of Physics and Astronomy, Johns Hopkins University, Baltimore, MD 21218, USA \\
$^{4}$Space Telescope Science Institute, Baltimore, MD 21218, USA \\
$^{5}$Department of Physics and Kavli Institute for Astrophysics and Space Research, Massachusetts Institute of Technology, Cambridge, MA 02139, USA \\
}

\date{Accepted 2024 October 07. Received 2024 August 30; in original form 2023 September 30}

\pubyear{2024}

\begin{document}
\label{firstpage}
\pagerange{\pageref{firstpage}--\pageref{lastpage}}
\maketitle

\begin{abstract}

As exoplanet surveys reach ever-higher sensitivities and durations, planets analogous to the solar system giant planets are increasingly within reach. HD~28185 is a Sun-like star known to host a $m\sin i=6~M_J$ planet on an Earth-like orbit; more recently, a brown dwarf with a more distant orbit has been claimed. In this work we present a comprehensive reanalysis of the HD 28185 system, based on 22 years of radial velocity observations and precision \textit{Hipparcos-Gaia} astrometry. We confirm the previous characterisation of HD~28185~b as a temperate giant planet, with its $385.92^{+0.06}_{-0.07}$~day orbital period giving it an Earth-like incident flux. In contrast, we substantially revise the parameters of HD~28185~c; with a new mass of $m=6.0\pm0.6~M_J$ we reclassify this companion as a super-jovian planet. HD~28185~c has an orbital period of $24.9^{+1.3}_{-1.1}$~years, a semi-major axis of $8.50^{+0.29}_{-0.26}$~AU, and a modest eccentricity of $0.15\pm0.04$, resulting in one of the most Saturn-like orbits of any known exoplanet. HD~28185~c lies at the current intersection of detection limits for RVs and direct imaging, and highlights how the discovery of giant planets at $\approx$10~AU separations is becoming increasingly routine.

\end{abstract}




\section{Introduction} \label{sec:intro}

As the search for exoplanets extends into its fourth decade, the population of giant planets at orbital scales comparable to those of the solar system ($a\gtrsim5$~AU) has come increasingly into focus. Though the most iconic extrasolar giant planets are the short-period Hot Jupiters, long-term radial velocity (RV) surveys have shown that giant planets are most frequently found at much more distant separations since at least the work of \citet{Cumming2008}. Historically the frequency of distant planets has been difficult to quantify due to the limited duration of RV surveys, but as observations slowly extend beyond 20-30~years for many stars it has become increasingly possible to detect giant planets with Jupiter-like and even Saturn-like orbits \citep[e.g.][]{Gregory2010, Rickman2019}. Recent estimates for the occurrence rate of planets more massive than Saturn in the period range $3000-10000$~d ($\approx$$3-10$~AU) include $6.9^{+4.2}_{-2.1}$ \citep{Wittenmyer2020} or $12.6^{+2.6}_{-2.0}$ giant planets per 100 stars \citep{Fulton2021}, demonstrating that these solar system-like giant planets are not infrequent components of exoplanetary systems. With this increasing clarity, the functional form of the distant giant planet occurrence rate has become a topic of debate; some studies have argued that giant planet occurrence declines beyond $a\gtrapprox3$~AU \citep{Fernandes2019, Fulton2021}, while others propose that the occurrence rate remains approximately constant in the same range \citep{Wittenmyer2020, Lagrange2023}.

Towards yet wider orbital separations ($\gtrapprox$10~AU), direct imaging is providing an increasingly clear picture of giant planet statistics. The most recent results from direct imaging surveys suggest that the occurrence rate of giant planet at separations of at $10<a<100$~AU is no higher than a few percent for FGK-type stars \citep{Nielsen2019, Vigan2021}. This appears to complement the hypothesis of decline in giant planet occurrence towards wider separations \citep{Fulton2021}. However, there remains an area of parameter space at orbital separations of $\approx$10~AU where the detection efficiency is low for both RVs and direct imaging. Further discoveries in this intersection of detection limits are therefore important for a complete picture of long-period giant planet occurrence from RVs and direct imaging.

An innovation that has significantly contributed to the study of long-period giant planets in recent years is \textit{Hipparcos-Gaia} astrometry. This is premised on the combination of proper motion data from the \textit{Hipparcos} and \textit{Gaia} missions \citep{Hipparcos, Gaia}, allowing for precise measurement of stellar tangential motions across the intervening $\sim$25~yr timespan \citep{Brandt2018, Kervella2019}. In the realm of exoplanet studies one of the key applications of this data is the characterisation of giant planets discovered through radial velocity surveys. RV planet detections suffer from the well-known $m\sin i$ degeneracy where the mass of the companion depends on the the orbital inclination $i$, which cannot be solved from RV data alone. \textit{Hipparcos-Gaia} astrometry has been used to break this degeneracy for a number of long-period RV planets \citep[e.g.][]{Li2021, BardalezGagliuffi2021, Venner2022, Philipot2023S}. In some cases this has shown that the orbits of these companions are viewed near to pole-on, revealing that certain RV planet candidates have masses in the brown dwarf or even stellar regime \citep{Venner2021}. Conversely, even for planets where the astrometric signal is too small to be detected, informative upper limits can be placed on the true mass allowing the companions to be confirmed as planets \citep{Errico2022}.

\thisstar{} is a Sun-like star long known to host a $m\sin i=6~M_J$ planet on an Earth-like orbit \citep{Santos2001}. The existence of an additional companion on a wider orbit has also been suspected, and this has recently manifested as a purported brown dwarf \citep{Rosenthal2021, Feng2022}. In this work we revisit the \thisstar{} system through a comprehensive analysis of published RV data and \textit{Hipparcos-Gaia} astrometry. We significantly revise the parameters of the outer companion in this system, finding that its mass has been substantially overestimated. \thisstarc{} is recovered here as a planet of super-jovian mass, with one of the longest orbital periods of any exoplanet detected (primarily) through RVs. This discovery expands our understanding of giant exoplanets at solar system-like scales, and highlights how a complete picture of giant planet occurrence rates is increasingly within reach.

\section{History of study} \label{sec:history}

The $V=7.8$~mag solar-type star \thisstar{} does not have a presence in the astronomical literature prior to the 1990s. The star was first identified as being relatively near to the Sun as a result of the \textit{Hipparcos} mission \citep{Hipparcos}, which measured a stellar parallax of $25.28\pm1.08$~mas. 
The search for exoplanets was beginning in earnest at this time, and \thisstar{} was included as a target for the CORALIE planet search \citep{Udry2000}. Subsequently, \citet{Santos2001} reported the discovery of a planetary companion to this star. They found that \thisstarb{} has a minimum mass of $5.7~M_J$ and has an orbit with a period of $383\pm2$~days and an eccentricity of $0.07\pm0.04$. \citet{Santos2001} considered the low orbital eccentricity of \thisstarb{} to be remarkable as most similar exoplanets known at the time had higher orbital eccentricities, and noted the similarity of the orbit of \thisstarb{} with that of Earth. Finally, the authors suggested there is evidence for a longer-period radial velocity signal, although they did not provide details.

Though no subsequent publication has reanalysed the CORALIE observations of \thisstar{}, in the context of a search for stellar companions to exoplanet host stars, \citet{Chauvin2006} mention the existence of an 11 \msyr{} radial velocity acceleration of \thisstar{} that suggests the existence of a second companion on a wider orbit. The source of this detection is not stated, but it does not originate from \citet{Santos2001}. The authors acquired an adaptive optics image of \thisstar{} using VLT/NACO \citep{Lenzen2003} and did not detect any companions, placing strong limits on any possible outer stellar companions in this system.

After the discovery of \thisstarb{} a number of other radial velocity surveys have observed the star. The following studies have published observations of \thisstar{}:

\begin{itemize}
	\item \citet{Minniti2009} presented an independent detection of \thisstarb{} from 3.6~years of observations with the MIKE spectrograph at the Magellan II telescope \citep{Bernstein2003}. The authors found a best-fit period of $379\pm2$~days and eccentricity of $0.05\pm0.03$ for the planet, similar to the results of \citet{Santos2001}, and determined a minimum mass of $6.7~M_J$. Finally, they noted that no radial velocity acceleration is evident in their data.
	\item \citet{Wittenmyer2009} observed \thisstar{} with the HET/HRS spectrograph \citep{Tull1998} for three seasons between 2004-2007 as part of an intensive search for new components in known planet-hosting systems. The authors did not find any evidence for additional planets around the star, particularly on short-period orbits where they were most sensitive. However, they did improve on the parameter precision for \thisstarb{} ($P=385.9\pm0.6$~days,\; $e=0.092\pm0.019$,\; $m\sin i=5.59\pm0.33~M_J$).
    \item \citet{Butler2017} published RV observations of \thisstar{} spanning $2004-2013$ from the Keck/HIRES planet search \citep{Vogt1994, Vogt2000}. As this study was primarily statistical in nature the amount of detail for this system is limited, but an orbital period of $P=383.72\pm0.67$~d and a semi-amplitude of $K=140.86\pm5.05$~\ms{} for \thisstarb{} are given. Additionally, their adopted model includes a $4.47\pm1.34$~\msyr{} RV acceleration, making this the first study to report an RV acceleration after \citet{Chauvin2006}.
    \item \citet{Rosenthal2021} provided an updated analysis of the Keck/HIRES RVs with an extension of the data to 2019. As with \citet{Butler2017} this study is primarily statistical and details are limited, but for \thisstarb{} the reported parameters are $a=1.045^{+0.016}_{-0.018}$~AU,\; $e=0.0629^{+0.0042}_{-0.0049}$, and $m\sin i=6.04\pm0.2~M_J$.
\end{itemize}

\citet{Rosenthal2021} additionally published the first orbital constraints on the outer companion as observed in the HIRES RVs. They report $a=15.9^{+7.3}_{-5.1}$~AU,\; $m\sin i=40^{+43}_{-28}~M_J$, and $e=0.26^{+0.12}_{-0.093}$. Based on the nominally high mass of this companion they classified as a brown dwarf; however, as their orbital coverage for this companion is highly incomplete, these parameters are highly uncertain and may not be robust.

Most recently, \citet{Feng2022} have presented a new solution for the \thisstar{} system as part of a large-scale joint analysis of RVs and astrometry for a number of systems. Their analysis is based on a selection of published RV datasets (chiefly MIKE and HIRES) plus previously unpublished PFS RVs \citep{Crane2010}, and astrometry from \textit{Hipparcos} and \textit{Gaia}. \citet{Feng2022} found similar parameters for \thisstarb{} as previous works. For the outer companion, which they name as \thisstarc{}, they report $P=17418^{+293}_{-673}$~d\; \big($=47.7^{+0.8}_{-1.8}$~yr; $a=13.18^{+0.52}_{-0.69}$~AU\big),\; $e=0.120^{+0.021}_{-0.022}$,\; $i=57.65^{+8.15}_{-5.75}\degree$,\; and $m=19.64^{+2.27}_{-2.14}~M_J$. This mass places \thisstarc{} well above the canonical deuterium-burning limit at 13~$M_J$, so this object would conventionally be classified as a brown dwarf.

\section{Stellar parameters} \label{sec:parameters}

\begin{table}
	\centering
	\caption{Observed and inferred parameters of \thisstar{}.}
	\label{table:parameters}
	\begin{tabular}{lrr}
		\hline
		Parameter & Value & Reference \\
		\hline
		R.A. $\alpha$ & 04:26:26.32 & \citetalias{GaiaEDR3} \\
		Declination $\delta$ & -10:33:02.95 & \citetalias{GaiaEDR3} \\
		Parallax $\varpi$ (mas) & $25.487\pm0.021$ & \citetalias{GaiaEDR3} \\
		Distance $d$ (pc) & $39.20\pm0.03$ & \citet{BailerJones2021} \\
		$V$ (mag) & $7.81\pm0.01$ & \citet{Tycho2} \\
		Spec. $T_{\text{eff}}$ (K) & $5621\pm22$ & \citet{Tsantaki2013} \\
		Spec. $\log g$ (cgs) & $4.36\pm0.05$ & \citet{Tsantaki2013} \\
		$\text{[Fe/H]}$ (dex) & $0.19\pm0.01$ & \citet{Tsantaki2013} \\
		\hline
		Model $T_{\text{eff}}$ (K) & $5602\pm36$ & This Work \\
		Luminosity ($L_\odot$) & $0.970\pm0.019$ & This work \\
		$M$ ($M_\odot$)\; $^{(a)}$ & $0.974\pm0.018$ & This Work \\
		$R$ ($R_\odot$) & $1.048\pm0.015$ & This Work \\
		Model $\log g$ (cgs) & $4.386\pm0.015$ & This Work \\
		Age $T$ (Gyr)\; $^{(a)}$ & $8.3\pm1.0$ & This Work \\
		\hline
		\multicolumn{3}{l}{$^{(a)}$ Includes systematic uncertainties (see text).}\\
	\end{tabular}
\end{table}

Owing to its status as a bright and nearby Sun-like star, and perhaps more significantly as a planet host, there is a rich body of study regarding the stellar parameters of \thisstar{} \citep[e.g.][]{Santos2004, Ghezzi2010, Ramirez2014, Maldonado2015, Soto2018}. Nonetheless, for the purposes of completeness we choose to retread this ground in this study.

\thisstar{} is a Sun-like star, with a spectroscopic classification of G5/6V \citep{Houk1999} or G6.5V-IV \citep{Gray2006}, at a distance of $39.20\pm0.03$~pc from the solar system \citep{BailerJones2021}. Its spectroscopic properties have been studied in detail based on HARPS data \citep[e.g.][]{Sousa2008, Tsantaki2013}. We list the basic observable properties of the star in Table~\ref{table:parameters}.

To derive the physical parameters of \thisstar{} we use the MIST isochrones \citep{Dotter2016, Choi2016}, constructing a model that uses \texttt{minimint} \citep{Koposov2021} for isochrone interpolation and \texttt{emcee} \citep{emcee} for posterior sampling. The data used for this model consists of photometry from Tycho-2 \citep{Tycho2}, \textit{Gaia}~EDR3 \citep{GaiaEDR3}, 2MASS \citep{2MASS}, and WISE \citep{WISE}; this data is reproduced in Table~\ref{table:photometry}. We supplement this with a prior on the distance (from the \textit{Gaia} parallax), and priors on $T_{\text{eff}}$ and [Fe/H] from \citet{Tsantaki2013}. However, the reported uncertainties on these spectroscopic parameters are strictly internal, and it is known that systematic uncertainties in parameters as found from comparison between different spectrographs may be larger than the nominal uncertainties \citep[e.g.][]{Brewer2016}. We thus conservatively double the widths of the adopted spectroscopic priors (i.e. $T_{\text{eff}}=5621\pm44$~K; $\text{[Fe/H]}=0.19\pm0.02$~dex). Due to the close distance of \thisstar{}, we assume there is no interstellar extinction in the photometry.

Our isochrone model returns a mass $M=0.974\pm0.014~M_\odot$ and radius $R=1.048\pm0.015~R_\odot$ for \thisstar{}. The resulting model $\log g$ of $4.386\pm0.015$ is in excellent agreement with the independent spectroscopic value \citep[$4.36\pm0.05$,][]{Tsantaki2013}, supporting the accuracy of these parameters. We find a posterior $T_{\text{eff}}$ of $5602\pm36$~K and a sub-solar luminosity $L=0.970\pm0.019~L_\odot$. The combination of all of these physical parameters leads to an isochronal age $T=8.3\pm0.9$~Gyr.

It is increasingly recognised that the uncertainties on stellar parameters are limited not just by measurement precision, but also by systematic differences between models. To attempt to account for this we use \texttt{kiauhoku} \citep{Tayar2022} to estimate the model uncertainties on the stellar mass and age. \texttt{kiauhoku} calculates the difference in physical parameters between four stellar models (MIST, YREC, Dartmouth, and GARSTEC) when assuming the same observables. For this exercise, we use our values of $T_{\text{eff}}$, $\log g$, and [Fe/H] as input variables.  \texttt{kiauhoku} returns large uncertainty terms of $\sigma_M=0.018~M_\odot$ and $\sigma_T=1.6$~Gyr. However, this is largely driven by the discrepant YREC parameters ($\Delta M_{\text{YREC}}=-0.030~M_\odot$, $\Delta T_{\text{YREC}}=3.4$~Gyr compared to the MIST values). The YREC models differ from the others in that they are calibrated to red giants rather than the Sun, and seeing as \thisstar{} is a close match for the Sun in most respects, it is perhaps unsurprising that the YREC models do not reproduce its parameters as well. Excluding the YREC models the remaining solar-calibrated models agree well, with $\sigma_M=0.007~M_\odot$ and $\sigma_T=0.3$~Gyr. We then conservatively increase these values by 50\%, i.e. $\sigma_M=0.010~M_\odot$, $\sigma_T=0.5$~Gyr. Adding these uncertainties in quadrature to our existing values, we adopt final values of $M=0.974\pm0.018~M_\odot$ and $T=8.3\pm1.0$~Gyr for the mass and age of \thisstar{}. We report our adopted stellar parameters in Table~\ref{table:parameters}. \thisstar{} is therefore a metal-rich star that is slightly lower in mass than the Sun but also substantially older and, as a result, is larger in size ($R=1.048\pm0.015~R_\odot$).

\begin{table}
	\centering
	\caption{Photometry of \thisstar{} used in the isochrone fit.}
	\label{table:photometry}
	\begin{tabular}{lrr}
		\hline
		Band & Magnitude (mag) & Reference \\
		\hline
		$B_\text{T}$ & $8.690\pm0.016$ & \citet{Tycho2} \\
		$V_\text{T}$ & $7.891\pm0.011$ & \citet{Tycho2} \\
		$G$ & $7.6400\pm0.0028$ & \citetalias{GaiaEDR3} \\
		$G_{\text{BP}}$ & $7.9881\pm0.0029$ & \citetalias{GaiaEDR3} \\
		$G_{\text{RP}}$ & $7.1235\pm0.0038$ & \citetalias{GaiaEDR3} \\
		$J$ & $6.578\pm0.026$ & \citet{2MASS} \\
		$H$ & $6.289\pm0.024$ & \citet{2MASS} \\
		$K_S$ & $6.185\pm0.029$ & \citet{2MASS} \\
		$W1$ & $6.164\pm0.045$ & \citet{WISE} \\
		$W2$ & $6.170\pm0.021$ & \citet{WISE} \\
		$W3$ & $6.184\pm0.015$ & \citet{WISE} \\
		$W4$ & $6.095\pm0.039$ & \citet{WISE} \\
		\hline
	\end{tabular}
\end{table}

There is no evidence that \thisstar{} has any stellar companions. \citet{Tokovinin2012} observed that UCAC4~399-005893 at a separation of 1600" has a similar proper motion to \thisstar{}, resulting in this pair being accessioned to the Washington Double Star Catalog \citep{WDS} as WDS~04264-1033. However, already in \citet{Tokovinin2012} this pair has a very low probability of physical association ($P_{\text{phys}}=0.001$), and with the addition of the \textit{Gaia} parallax it can now be determined that UCAC4~399-005893 lies at a distance of $107$~pc \citep{BailerJones2021}, hence far in the background to \thisstar{}. No star found in \textit{Gaia}~DR3 can be identified as companions to \thisstar{} due to a lack of agreement in proper motion and distance. Additionally, adaptive optics imaging from \citet{Chauvin2006} excludes much of the possible parameter space for stellar companions down to separations of less than an arcsecond. \thisstar{} therefore appears to be a single star.

\section{Method} \label{sec:method}

\subsection{Data}  \label{subsec:data}

We have collated RV data for \thisstar{} from the published literature. A total of 7 separate datasets for 6 instruments are represented in the radial velocity data:

\begin{itemize}
    \item 40 CORALIE RVs spanning $1999-2001$, the discovery data for \thisstarb{}, are taken from \citet{Santos2001}. These observations do not overlap temporally with any of the remaining datasets.
    \item 16 MIKE RVs spanning $2002-2009$ are from \citet{Feng2022}. Compared to the MIKE data published in \citet{Minniti2009}, this includes one additional observation.
    \item 34 HRS RVs spanning $2005-2007$ are taken from \citet{Wittenmyer2009}. These observations were taken as part of a targeted survey of known planet hosts; after the conclusion of this survey, observations of \thisstar{} were not continued (B. Bowler, private communication).
    \item 34 HIRES RVs spanning $2004-2019$ were published in \citet{Rosenthal2021}. Four of these observations precede the 2004 instrument upgrade, and we observe that these observations show a large offset compared to the post-upgrade RVs; we thus treat these as two separate datasets for our model. We additionally opt to take nightly medians of the RVs. This results in a final set of 25 (4 + 21) HIRES RVs.
    \item 29 PFS RVs spanning $2011-2022$ were published in \citet{Feng2022}. We take nightly bins of this data, resulting in a reduced set of 22 RV observations.
    \item Finally, a limited set of 10 RVs from the HARPS spectrograph \citep{Mayor2003} spanning $2003-2004$ are available in the ESO science archive,\footnote{\url{https://archive.eso.org/wdb/wdb/adp/phase3_spectral/form?phase3_collection=HARPS}} and have been published in re-reduced form by \citet{Trifonov2020}. Of these, we omit one relatively low-S/N RV ($\text{BJD}=2453263.9$) as it is a significant ($>$5$\sigma$) outlier compared to the remaining measurements. This results in a set of 9 HARPS RVs used in this work.\footnote{We note that there are two additional HARPS observations from 2011 in the ESO archive, but these were collected under different instrument settings to the aforementioned observations and show a significant RV offset, rendering them impertinent to the present analysis. Furthermore, a rich set of HARPS RVs spanning $2018-2021$ can also be found in the ESO archive; however, we abstain from analysis of this data prior to its formal publication.}
\end{itemize}

In summary, our adopted RV data for \thisstar{} consist of 146 measurements spread across 7 separate datasets, with a first-to-last time duration of 22.3~yr.

\begin{table}
	\centering
	\caption{Proper motions of \thisstar{} from the \textit{Hipparcos-Gaia} Catalog of Accelerations \citep{Brandt2021}.}
	\label{table:astrometry}
	\begin{tabular}{lcrr}
		\hline
		Measurement &  & $\mu_{\alpha}$ (\masyr{})  & $\mu_{\delta}$ (\masyr{}) \\
		\hline
		\textit{Hipparcos} & $\mu_{\text{H}}$ & $+82.967\pm0.923$ & $-58.736\pm0.801$ \\
		\textit{Gaia} EDR3 & $\mu_{\text{G}}$ & $+84.070\pm0.024$ & $-59.637\pm0.021$ \\
		\textit{Hipparcos-Gaia} & $\mu_{\text{HG}}$ & $+84.181\pm0.024$ & $-59.529\pm0.019$ \\
		\hline
	\end{tabular}
\end{table}

The astrometric data used in this work comes from the \textit{Hipparcos-Gaia} Catalog of Accelerations \citep[HGCA;][]{Brandt2021}. We refer the reader to \citet{Brandt2018, Brandt2021} and further to \citet{Brandt2019} for a full description of this data, but as a brief summary, the astrometric data in the HGCA consist of three \textit{de facto} independent measurements of stellar proper motion; these are the \textit{Hipparcos} proper motion ($\mu_{\text{H}}$), here a linear combination of the original and new \textit{Hipparcos} reductions \citep{Hipparcos, HipparcosNew}, measured at an approximate epoch of 1991.25; the \textit{Gaia} proper motion ($\mu_{\text{G}}$), here from \textit{Gaia}~EDR3 \citep{GaiaEDR3}, measured at approximately epoch 2016.0; and lastly the \textit{Hipparcos-Gaia} mean proper motion ($\mu_{\text{HG}}$), which is derived from the change in sky position measured by the instruments and is hence time-averaged between the two epochs.

We reproduce the astrometry of \thisstar{} in the HGCA in Table~\ref{table:astrometry}. As is typical for this data the \textit{Hipparcos} proper astrometry has the largest uncertainties, but the \textit{Gaia} and \textit{Hipparcos-Gaia} proper motions are highly precise (0.025~\masyr{}~$\approx$~5~\ms{} at the distance of \thisstar{}) and are therefore strongly sensitive to long-term deviations in the motion of this star. For \thisstar{} the $\mu_{\text{G}}-\mu_{\text{HG}}$ proper motion anomaly is ($-0.111\pm0.034,-0.108\pm0.028$~\masyr{}), a $>$4$\sigma$ significant difference \citep[$\chi^2=22.9$,][]{Brandt2021} equivalent to a net velocity change of $\sim$30~\ms{}.

\subsection{Model}

\subsubsection{Model 1} \label{subsec:model1}

\begin{table*}
	\centering
	\caption{Posterior parameters for our two joint fits to the RVs and astrometry. All values are medians and 1$\sigma$ confidence intervals.}
	\label{table:results}
	\begin{tabular}{lcccccc}
		\hline
		Parameter & \multicolumn{3}{c}{Model 1 (adopted)} & \multicolumn{3}{c}{Model 2} \\
		  & \thisstarb{} & \multicolumn{2}{c}{\thisstarc{}} & \thisstarb{} & \multicolumn{2}{c}{\thisstarc{}}\\
		\hline
		Period $P$ (days) & $385.92^{+0.06}_{-0.07}$ & \multicolumn{2}{c}{$9090^{+460}_{-390}$} & $385.96\pm0.06$ & \multicolumn{2}{c}{$9170^{+400}_{-330}$} \\
		Period $P$ (years) & $1.0566\pm0.0002$ & \multicolumn{2}{c}{$24.9^{+1.3}_{-1.1}$} & $1.0567\pm0.0002$ & \multicolumn{2}{c}{$25.1^{+1.1}_{-0.9}$} \\
		RV semi-amplitude $K$ (\ms{}) & $164.8\pm0.9$ & \multicolumn{2}{c}{$53.3^{+5.1}_{-4.7}$} & $164.4\pm1.0$ & \multicolumn{2}{c}{$54.2^{+4.2}_{-4.0}$} \\
		Eccentricity $e$ & $0.063\pm0.004$ & \multicolumn{2}{c}{$0.15\pm0.04$} & $0.063\pm0.004$ & \multicolumn{2}{c}{$0.14^{+0.04}_{-0.05}$} \\
		Argument of periastron $\omega$ ($\degree$) & $355.1\pm3.9$ & \multicolumn{2}{c}{$162\pm8$} & $354\pm4$ & \multicolumn{2}{c}{$163^{+9}_{-8}$} \\
		Time of periastron $T_{\text{P}}$ (JD) & $2451870.2\pm4.5$ & \multicolumn{2}{c}{$2460790^{+350}_{-280}$} & $2451869.0\pm4.3$ & \multicolumn{2}{c}{$2460870^{+370}_{-290}$} \\
		Minimum mass $m\sin i$ ($M_J$) & $5.85\pm0.08$ & \multicolumn{2}{c}{$5.4^{+0.6}_{-0.5}$} & $5.80\pm0.08$ & \multicolumn{2}{c}{$5.5\pm0.4$} \\
		Semi-major axis $a$ (AU) & $1.034\pm0.006$ & \multicolumn{2}{c}{$8.50^{+0.29}_{-0.26}$} & $1.031\pm0.006$ & \multicolumn{2}{c}{$8.52^{+0.25}_{-0.22}$} \\
		\hline
		Orbital inclination $i$ ($\degree$) & -- & $66^{+11}_{-9}$ & $114^{+9}_{-11}$ & -- & $71\pm12$ & $109\pm12$ \\
		Longitude of node $\Omega$ ($\degree$) & -- & $271^{+15}_{-21}$ & $178^{+18}_{-14}$ & -- & $270^{+30}_{-35}$ & $187^{+32}_{-33}$ \\
		$\sin i$ & -- & \multicolumn{2}{c}{$0.91^{+0.06}_{-0.07}$} & -- & \multicolumn{2}{c}{$0.95^{+0.05}_{-0.09}$} \\
		Orbital velocity semi-amplitude $\frac{K}{\sin i}$ (\ms{}) & -- & \multicolumn{2}{c}{$59.0^{+5.7}_{-5.2}$} & -- & \multicolumn{2}{c}{$58.6^{+4.8}_{-4.3}$} \\
		Mass $m$ ($M_J$) & -- & \multicolumn{2}{c}{$6.0\pm0.6$} & -- & \multicolumn{2}{c}{$5.9^{+0.5}_{-0.4}$} \\
		\hline
		Barycentre R.A. P.M. $\mu_{\alpha\text{,bary}}$ (\masyr{}) & \multicolumn{3}{c}{$+84.172\pm0.026$} & -- & \multicolumn{2}{c}{--} \\
		Barycentre dec. P.M. $\mu_{\delta\text{,bary}}$ (\masyr{}) & \multicolumn{3}{c}{$-59.534\pm0.020$} & -- & \multicolumn{2}{c}{--} \\
		\hline
		CORALIE RV offset (\ms{}) & \multicolumn{3}{c}{$50305.9\pm8.2$} & \multicolumn{3}{c}{50311.5 $^{(a)}$} \\
		MIKE RV offset (\ms{}) & \multicolumn{3}{c}{$55.3^{+4.6}_{-4.9}$} & \multicolumn{3}{c}{54.8 $^{(a)}$} \\
		PFS RV offset (\ms{}) & \multicolumn{3}{c}{$-31.5^{+4.0}_{-4.2}$} & \multicolumn{3}{c}{-32.9 $^{(a)}$} \\
		HRS RV offset (\ms{}) & \multicolumn{3}{c}{$95.1^{+4.3}_{-4.4}$} & \multicolumn{3}{c}{93.3 $^{(a)}$} \\
		HARPS RV offset (\ms{}) & \multicolumn{3}{c}{$75.6^{+4.0}_{-4.3}$} & \multicolumn{3}{c}{--} \\
		HIRES pre-upgrade RV offset (\ms{}) & \multicolumn{3}{c}{$-118.3^{+6.0}_{-6.6}$} & \multicolumn{3}{c}{--} \\
		HIRES post-upgrade RV offset (\ms{}) & \multicolumn{3}{c}{$20.5^{+3.8}_{-4.1}$} & \multicolumn{3}{c}{18.8 $^{(a)}$} \\
		\hline
		CORALIE RV jitter (\ms{}) & \multicolumn{3}{c}{$9.0^{+2.1}_{-1.8}$} & \multicolumn{3}{c}{$8.0^{+1.9}_{-1.7}$} \\
		MIKE RV jitter (\ms{}) & \multicolumn{3}{c}{$12.2^{+3.4}_{-2.6}$} & \multicolumn{3}{c}{$12.3^{+3.4}_{-2.6}$} \\
		PFS RV jitter (\ms{}) & \multicolumn{3}{c}{$4.1^{+1.0}_{-0.7}$} & \multicolumn{3}{c}{$4.2^{+1.0}_{-0.8}$} \\
		HRS RV jitter (\ms{}) & \multicolumn{3}{c}{$1.8^{+1.7}_{-1.2}$} & \multicolumn{3}{c}{$0.0^{+0.4}_{-0.0}$} \\
		HARPS RV jitter (\ms{}) & \multicolumn{3}{c}{$6.0^{+2.3}_{-1.6}$} & \multicolumn{3}{c}{--} \\
		HIRES pre-upgrade RV jitter (\ms{}) & \multicolumn{3}{c}{$7.7^{+9.6}_{-4.2}$} & \multicolumn{3}{c}{--} \\
		HIRES post-upgrade RV jitter (\ms{}) & \multicolumn{3}{c}{$2.2^{+0.7}_{-0.5}$} & \multicolumn{3}{c}{$2.3^{+0.7}_{-0.5}$} \\
		\hline
        \multicolumn{7}{l}{$^{(a)}$ \texttt{orvara} normalises over RV zero-points, so only the best-fit values are reported \citep{orvara}.} \\
	\end{tabular}
\end{table*}

\begin{figure*}
	\includegraphics[width=\textwidth]{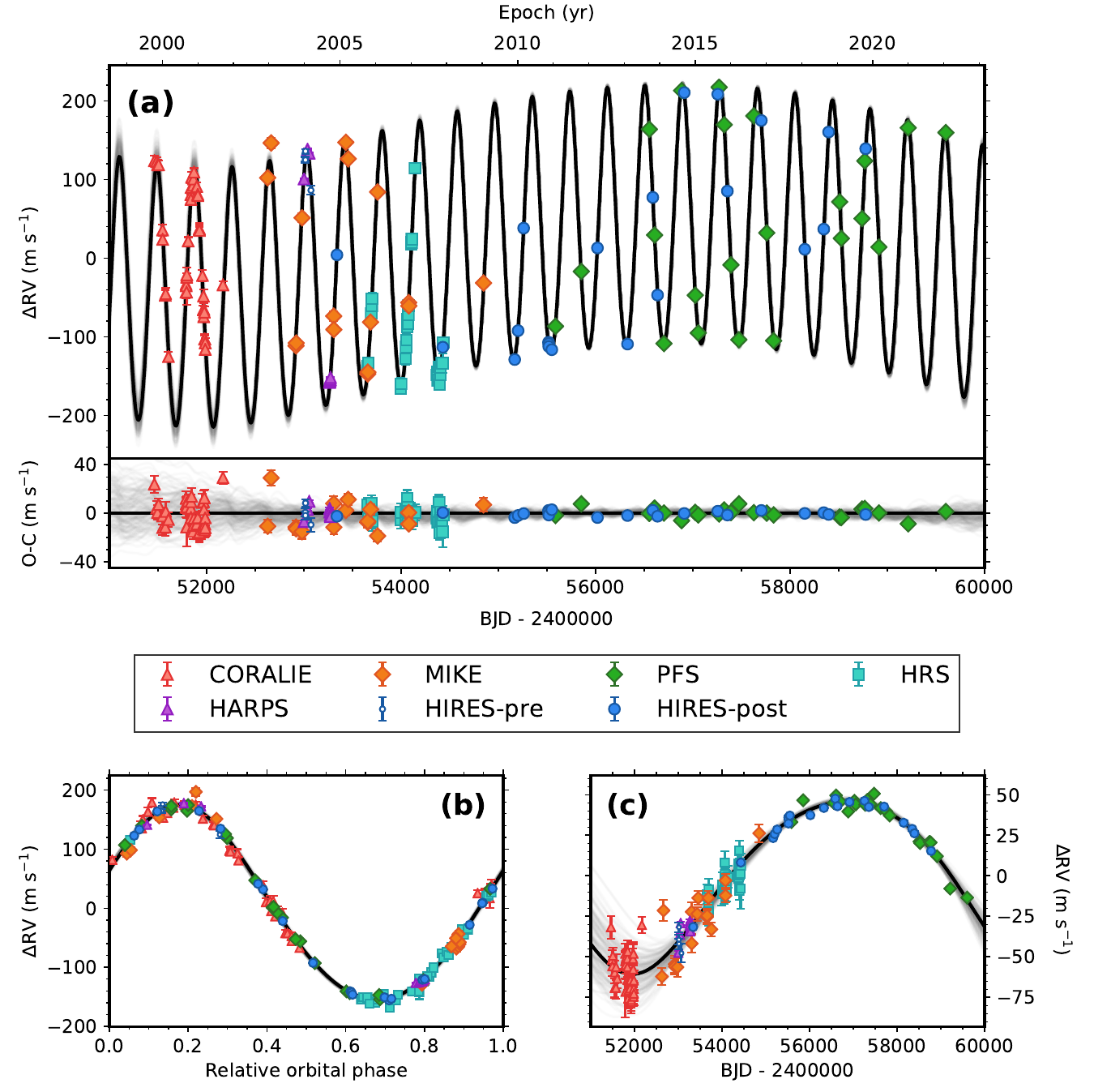}
	\caption{Our fit to the radial velocity of \thisstar{} from Model~1. Panel (a) shows the observed radial velocities over time; panel (b) shows the variation caused by \thisstarb{} phase-folded to its orbital period, with the signal of \thisstarc{} subtracted; and panel (c) shows the RV variation caused by \thisstarc{} with \thisstarb{} subtracted. Even with the combination of 7 RV datasets from 6 different instruments, the orbit of \thisstarc{} is not completely covered by the 22-year timespan of the RVs; resolution of its orbit therefore depends crucially on the extension in observational duration offered by the astrometry (see Figure~\ref{figure:astrometry})}
	\label{figure:RV}
\end{figure*}

\begin{figure*}
	\includegraphics[width=\textwidth]{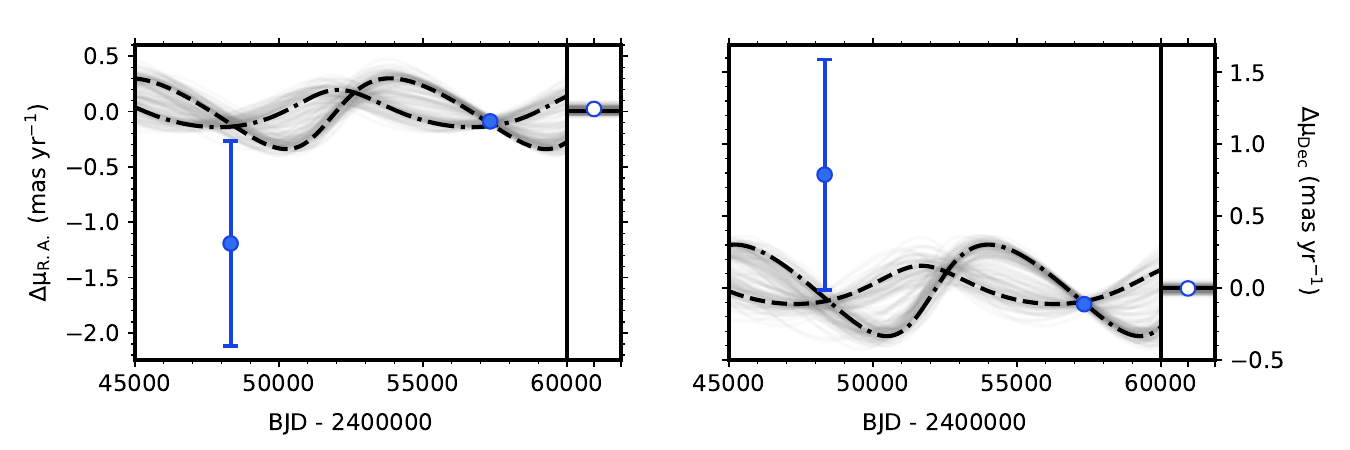}
	\caption{Our fit to the proper motion variability of \thisstar{} caused by \thisstarc{} from Model~1, in right ascension (\textit{left}) and declination (\textit{right})\bmaroon{, normalised to the barycentre proper motion ($\mu_{\alpha,\text{bary}}$, $\mu_{\delta,\text{bary}}$)}. The two filled points in the main panels are the \textit{Hipparcos} and \textit{Gaia} proper motions while the unfilled points in the side panels represent the \textit{Hipparcos-Gaia} mean proper motions. The detection of \thisstarc{} in the astrometry is driven by the highly precise \textit{Gaia} and \textit{Hipparcos-Gaia} proper motions; in concert with the RV data, the comparatively small net acceleration allows us to resolve the orbital period of \thisstarc{}. However, due to the low precision of the \textit{Hipparcos} proper motion there are two families of orbital solutions for \thisstarc{}, represented by the dotted and dash-dotted lines; see text and Figure~\ref{figure:bimodal} for elaboration.}
	\label{figure:astrometry}
\end{figure*}

The primary model used in this work to fit the radial velocities and astrometry of \thisstar{} is based on the one in \citet{Venner2021}, and we refer to reader to that work for a detailed description of the method. \bmaroon{In this model, we use 30 variable parameters to fit the data described in Section~\ref{subsec:data}. Two reflect the stellar mass $M_*$ and the parallax $\varpi$, which are assigned Gaussian priors equal to the values given in Table~\ref{table:parameters}. Five define the orbit of \thisstarb{} as required to fit the RVs, these being the orbital period $P$, the RV semi-amplitude $K$, the eccentricity $e$ and argument of periastron $\omega$ (parameterised as $\sqrt{e}\sin\omega$ and $\sqrt{e}\cos\omega$), and the time of periastron $T_{\text{P}}$. Seven parameters then describe the orbit of \thisstarc{}, of which the first five are analogous to those previously listed for \thisstarb{} and the remainder are the orbital inclination $i$ and the longitude of node $\Omega$ as required to fit to the astrometry. Finally, there are two normalising terms for the proper motion of the system barycentre ($\mu_{\alpha\text{,bary}}$ and $\mu_{\delta\text{,bary}}$), and a total of 14 parameters delineate the offsets and jitter terms necessary for each of the seven RV datasets.}

\bmaroon{Compared to previous iterations of this model \citep{Venner2021, Venner2022}, the main alterations made for this work} concern the inclusion of a second companion in the fit. \bmaroon{Here we have made the explicit assumption that \thisstarb{} does not contribute to the astrometry.} The astrometric signal of \thisstarb{} is expected to be relatively large \citep[$\sim$0.15~mas following][equation~6, assuming $m=m\sin i$]{Sozzetti2005}, but as its orbital period is shorter that the $\approx$3-year observing spans of \textit{Hipparcos} and \textit{Gaia}~DR3 its reflex signal in the \textit{Hipparcos-Gaia} astrometry will be averaged out and reduced in amplitude \citep{Kervella2019}. More critical, however, is the $1.06$~year orbital period of \thisstarb{}; it is exceptionally difficult to detect orbits with $P\approx1$~yr using astrometry because the reflex signal is liable to absorption by the parallax \citep[compare the paucity of \textit{Gaia}~DR3 astrometric orbits with $\approx$1~yr periods in][figure~3]{Halbwachs2023}. This means that it is likely that the astrometric signal of \thisstarb{} has already been attenuated or altogether removed from the \textit{Hipparcos-Gaia} proper motions as a result of the processing of the component astrometric solutions.

We use the Markov Chain Monte Carlo sampler \texttt{emcee} \citep{emcee} the explore the posterior parameter space. 60 walkers were used to sample the model for $5\times10^5$ steps, with confirmation of convergence of the MCMC performed as in \citet{Venner2021}. We then discarded the first half of the chain as burn-in and saved every $150^{\text{th}}$ step to extract our final posteriors.

\subsubsection[Model 2: orvara]{Model 2: \texttt{orvara}} \label{subsec:model2}

To independently test the accuracy of the Model~1, we also perform a joint fit to the radial velocities and astrometry of \thisstar{} using \texttt{orvara} \citep{orvara}. \texttt{orvara} is designed for fitting any combination of RV, imaging, and \textit{Hipparcos-Gaia} astrometry data for stars with one or more orbiting companions, and has previously been used to jointly fit orbits of a number of exoplanets with data from RVs and \textit{Hipparcos-Gaia} astrometry \citep[e.g.][]{Li2021}.

A salient difference compared to Model~1 is that the formalism of \texttt{orvara} requires fitting of astrometric reflex signals for all companions; this means that \thisstarb{} is fitted for in the \textit{Hipparcos-Gaia} astrometry, and hence two additional terms for the orbital inclination $i$ and longitude of node $\Omega$ for \thisstarb{} are included in the model. \bmaroon{However, as argued previously in Section~\ref{subsec:model1}, the astrometric signal of \thisstarb{} is unlikely to be detectable.} Fitting four variables to three measurements of proper motion \bmaroon{also} hampers convergence. As a simplistic intervention, we assign a prior on the mass of \thisstarb{} forcing $m\approx m\sin i$ and then disregard its astrometric posteriors. This reduces the impact of \thisstarb{} on the astrometric fit, but slight differences may still remain compared to Model~1.

For Model~2 we again use the data described in Section~\ref{subsec:data}, however due to computational limitations we exclude the small HARPS and pre-upgrade HIRES RV datasets. We use the parallel-tempered MCMC sampler \texttt{ptemcee} \citep{Vousden2016, ptemcee} to explore the parameter space. We use 100 walkers and 20 temperatures and sample the model for $2\times10^5$ steps, discarding the first quarter of the chain as burn-in and saving at every $50^{\text{th}}$ step for the final posteriors.

\section{Results} \label{sec:results}

The results of our two joint fits to the radial velocities and astrometry of \thisstar{} are shown in Table~\ref{table:results}. We describe the results of both models in turn below.

\subsection{Results for Model 1} \label{subsec:results_model1}

In Figure~\ref{figure:RV} we present the RV half of our joint fit from Model~1. Beginning with the parameters of \thisstarb{}, we recover a tightly constrained orbit from the RV data; the key observable parameters are $P=385.92^{+0.06}_{-0.07}$~days, $K=164.8\pm0.9$~\ms{}, and $e=0.063\pm0.004$. Though the proximity of the orbital period of \thisstarb{} to one year ($P=1.0566\pm0.0002$~yr) means that the entire orbit cannot be covered in any single season, the 22-year timespan of the RV observations means that most of the planetary orbit has now been sampled as a result of the slow drift in the orbital phase that can be observed from Earth. In combination with our adopted stellar mass of $M=0.974\pm0.018~M_\odot$, the observable parameters for \thisstarb{} result in a super-jovian minimum mass of $m\sin i=5.85\pm0.08~M_J$.

Along with the reflex signal from \thisstarb{}, a long-term acceleration is clearly visible in Figure~\ref{figure:RV}. We note at this point that in the study of similar long-term RV signals a perennial concern is represented by solar-like magnetic cycles, which frequently act on timescales of decades and can produce planet-like RV signals \citep[e.g.][]{Endl2016, Diaz2016}. This can, however, be discarded as an explanation for this RV signal as the HIRES spectra show no significant variability in the S-index, a stellar activity proxy which is highly sensitive to magnetic cycles \citep{Butler2017, Rosenthal2021}. The long-term RV signal must therefore correspond to an outer companion, as has been inferred in previous publications.

Unfortunately, the orbit of this outer companion is not completely sampled by the extant RV observations. Furthermore, the fiducial RV minimum is covered only by the CORALIE data, which does not overlap with the remaining datasets; this means that the offset between CORALIE and the remaining RV data can be only loosely constrained when considered directly, and it is hence not possible to confidently distinguish the long-term acceleration from a quadratic trend from the RVs alone. Resolution of orbit of \thisstarc{} therefore depends crucially on the astrometry.

In Figure~\ref{figure:astrometry} we plot our fit to the \textit{Hipparcos-Gaia} astrometry of \thisstar{}. As noted in Section~\ref{subsec:data}, of the three proper motion measurements, the \textit{Hipparcos} proper motion is too imprecise to significantly contribute to the fit. Almost all of the astrometric information therefore comes from the difference between the \textit{Gaia} proper motion and the time-averaged \textit{Hipparcos-Gaia} proper motion. Though the measured velocity change between these measurements is significant ($\sim$30~\ms{}, Section~\ref{subsec:data}), \bmaroon{this} is small compared to the $\approx$100~\ms{} range of variability observed in the RVs; in other words, the \textit{net} tangential velocity change over the astrometric timespan is smaller than the \textit{total} radial velocity change over the RV timespan. As the tangential velocity semi-amplitude is equivalent to $\frac{K}{\sin i}$ \citep[i.e. $\geq$\textit{K};][]{Venner2021}, \bmaroon{the total change in tangential velocity cannot be smaller than the corresponding change in radial velocity; however, the \textit{net} change can be smaller if the period is close to $\sim 25$~yr interval between the \textit{Hipparcos} and \textit{Gaia}~DR3 epochs, as the signal will be averaged out from the mean proper motion}. and the \textit{Hipparcos-Gaia} proper motion represents the integrated motion across in the $\sim$25~year interval between the \textit{Hipparcos} and \textit{Gaia}~DR3 epochs, \bmaroon{We therefore} infer that the only possible explanation for the comparatively small size of the \bmaroon{net astrometric signal} is that there was a reversal in acceleration before the beginning of RV observations (separate to the RV reversal around $\sim$2015 present in the HIRES and PFS data); this is in turn best explained by hypothesising that \thisstarc{} completed about one complete orbit in the $\sim$25~year interval between \textit{Hipparcos} and \textit{Gaia} observations.

\begin{figure}
    \includegraphics[width=\columnwidth,clip=true,trim=0.1cm 0.5cm 0.1cm 0.5cm]{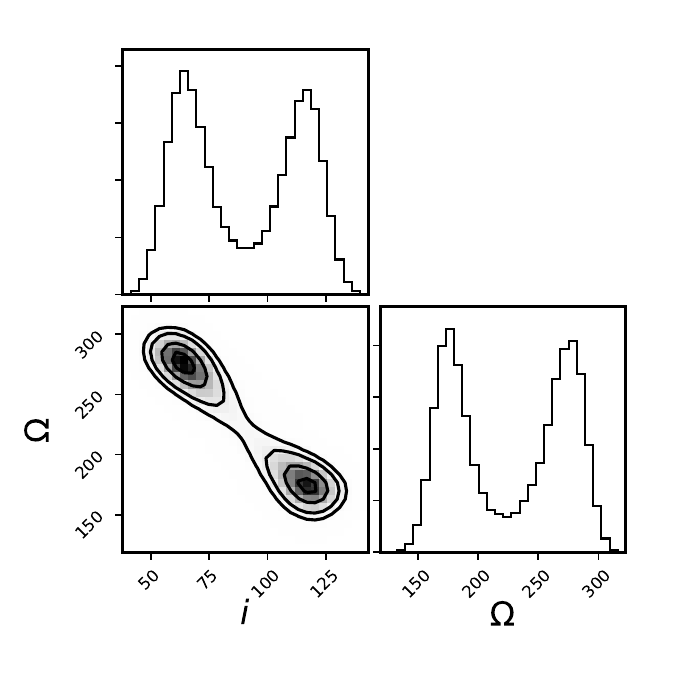}
    \caption{A \texttt{corner} plot \citep{corner} of orbital inclination $i$ versus longitude of node $\Omega$ for \thisstarc{}. Due to limitations of the astrometric data there are two peaks in the posterior which are mirrored around $i=$~90$\degree$, i.e. $i=66^{+11}_{-9}\degree$, $\Omega=271^{+15}_{-21}\degree$, and $i=114^{+9}_{-11}\degree$, $\Omega=178^{+18}_{-14}\degree$.}
    \label{figure:bimodal}
\end{figure}

\begin{table*}
	\centering
	\caption{Comparison of published parameters for \thisstarb{}.}
	\label{table:planet_b_comparison}
    \begin{tabular}{lccccc}
        \hline
        Parameter & \citet{Santos2001} & \citet{Minniti2009} & \citet{Wittenmyer2009} & \citet{Feng2022} & This work (Model 1) \\
        \hline
        $P$ (days) & $383\pm2$ & $379\pm2$ & $385.9\pm0.6$ & $385.528^{+0.044}_{-0.055}$ & $385.92^{+0.06}_{-0.07}$ \\
        $K$ (\ms{}) & $161\pm11$ & $163.5\pm3$ & $158.8\pm4.2$ & $163.66^{+0.65}_{-0.53}$ & $164.8\pm0.9$ \\
        $e$ & $0.07\pm0.04$ & $0.05\pm0.03$ & $0.092\pm0.019$ & $0.055^{+0.004}_{-0.003}$ & $0.063\pm0.004$ \\
        $\omega$ (degrees) & $351\pm25$ & $44\pm2$ & $351.9\pm8.2$ & $356.60^{+3.50}_{-3.16}$ & $355.1\pm3.9$ \\
        $m\sin i$ ($M_J$) & 5.7 & 6.7 & $5.59\pm0.33$ & $5.84^{+0.51}_{-0.49}$ & $5.85\pm0.08$ \\
        $a$ (AU) & 1.03 & 1.1 & $1.032\pm0.019$ & -- & $1.034\pm0.006$ \\
        \hline
    \end{tabular}
\end{table*}

This hypothesis is strongly borne out by our joint fit, from which we measure a well-constrained orbital period of $9090^{+460}_{-390}$~d \big($24.9^{+1.3}_{-1.1}$~yr\big) for \thisstarc{}. With the period resolved, it is thus possible to confidently constrain $K$ from the RV data ($53.3^{+5.1}_{-4.7}$~\ms{}). Additionally, though its precision is limited by the incompleteness of the RV coverage, we are able to constrain the orbital eccentricity with reasonably high confidence ($e=0.15\pm0.04$). This results in a minimum mass of $m\sin i=5.4^{+0.6}_{-0.5}~M_J$ for \thisstarc{}. Though the astrometric data allows us to constrain the orbital inclination of this outer companion, the interpretation of this is complicated by the existence of a degeneracy in $i$ which is symmetric around $90\degree$, i.e. $i=66^{+11}_{-9}\degree$ or $114^{+9}_{-11}\degree$. Similar degeneracies in orbital inclination have frequently been found in joint fits of \textit{Hipparcos-Gaia} astrometry and RVs for exoplanets \citep[e.g.][]{Li2021, Xiao2023, Philipot2023L}, and occur in the circumstance where the \textit{Hipparcos} proper motion is not sufficiently precise to independently detect the predicted variation in stellar tangential motion. When this is the case, the amplitude of tangential motion can still be constrained from the difference between the \textit{Gaia} and \textit{Hipparcos-Gaia} proper motions, but as this constitutes only two measurements and the latter is not time-resolved, in this scenario it is not possible to uniquely solve for both $i$ and the longitude of node $\Omega$ (i.e. the \textit{direction} of motion is ambiguous \bmaroon{between clockwise and anticlockwise}).

We demonstrate this covariance between $i_\text{c}$ and $\Omega_\text{c}$ as it occurs in our joint model in Figure~\ref{figure:bimodal}; meanwhile, in Figure~\ref{figure:astrometry} this degeneracy manifests in the symmetry of the best-fit solutions between the two planes of tangential motion. While we cannot uniquely solve for the orbital inclination of \thisstarc{}, because the two modes are symmetric around $90\degree$ we can solve for the value of $\sin i$ \big($0.91^{+0.06}_{-0.07}$\big). This therefore allows for a unique value for the true mass of $m=6.0\pm0.6~M_J$ for \thisstarc{}. We therefore determine that \thisstarc{} is a planet of super-jovian mass.

Based on the fitted orbital periods and our adopted stellar mass of $M=0.974\pm0.018~M_\odot$, we derive semi-major axes of $1.034\pm0.006$~AU and $8.50^{+0.29}_{-0.26}$~AU for \thisstarb{} and~c respectively. Combined with their modest eccentricities, the orbits of these two planets are comparable to those of Earth and Saturn among the solar system planets. We return to further consider their properties in Section~\ref{sec:discussion}.

\subsection{Results for Model 2} \label{subsec:results_model2}

The results from Model~2 are in very good agreement with those from Model~1, so it will be sufficient to discuss these briefly.

All planetary parameters which can be compared between the models are consistent at the 1$\sigma$ level. For the main observable parameters, we find $P_{\text{b}}=385.96\pm0.06$~d, $K_{\text{b}}=164.4\pm1.0$~\ms{}, and $e_{\text{b}}=0.063\pm0.004$ for \thisstarb{} and $P_{\text{c}}=9180^{+400}_{-340}$~d ($25.1^{+1.1}_{-0.9}$~yr), $K_{\text{c}}=54.2^{+4.2}_{-4.0}$~\ms{}, and $e_{\text{c}}=0.14^{+0.04}_{-0.05}$ for \thisstarc{}, all nearly identical to Model~1. The resulting semi-major axes for the two planets are $1.031\pm0.006$~AU and $8.52^{+0.25}_{-0.22}$~AU and their minimum masses are $5.80\pm0.08~M_J$ and $5.5\pm0.4~M_J$ respectively.

Though the inclusion of \thisstarb{} to the astrometric fit introduces a degree of ``fuzziness" to the relevant parameters for \thisstarc{}, they still agree very well with Model~1. Bimodality for $i_{\text{c}}$ and $\Omega_{\text{c}}$ is less pronounced but we again find two main solutions, the first with $i=71\pm12\degree$, $\Omega=270^{+30}_{-35}\degree$ and the second with $i=109\pm12\degree$, $\Omega=187^{+32}_{-33}\degree$. The corresponding true mass of \thisstarc{} is $m=5.9^{+0.5}_{-0.4}~M_J$. All of these posteriors agree extremely well with Model~1.

The strong agreement between our two models supports the validity of our results. For the remainder of this work we adopt the results from Model~1, but our interpretation would not change significantly if the results of Model~2 were used instead. Finally, we provide a selection of figures from Model~2 in Appendix~\ref{appendix:orvara}.

\section{Discussion} \label{sec:discussion}

\subsection{HD 28185 b, the archetypal temperate Jupiter} \label{subsec:planet_b}


At the time of its discovery \thisstarb{} was the first known exoplanet with an Earth-like orbit, with $P=383\pm2$~d, $e=0.07\pm0.04$, and $m\sin i=5.7~M_J$ \citep{Santos2001}. This original characterisation of \thisstarb{} has been resoundingly confirmed through the two subsequent decades of continued RV observations, as we summarise in Table~\ref{table:planet_b_comparison} (and see further in Section~\ref{sec:history}). In this work we continue this with a highly precise characterisation of this planet's parameters, with $P=385.92^{+0.06}_{-0.07}$~d, $e=0.063\pm0.004$, and $m\sin i=5.85\pm0.08~M_J$.

\thisstarb{} is the archetype of a now-numerous class of temperate giant planets. RV surveys have find a $5-10$\% occurrence rate for giant planets at Earth-like orbital separations around solar-type stars \citep{Wittenmyer2020, Fulton2021}. This approximates the location of the circumstellar habitable zone, for which the occurrence of transiting giant planet candidates was found to be comparable by \citet{Hill2018}.

\subsection{HD 28185 c} \label{subsec:planet_c}

\begin{table*}
	\centering
	\caption{Parameter comparison for \thisstarc{}.}
	\label{table:planet_c_comparison}
    \begin{tabular}{lllll}
        \hline
        Parameter & \citet{Rosenthal2021} & \citet{Feng2022} & Model 1 (adopted) & Model 2 \\
        \hline
        $P$ (days) & -- & $17418^{+293}_{-673}$ & $9090^{+460}_{-390}$ & $9180^{+400}_{-340}$ \\
        $a$ (AU) & $15.9^{+7.3}_{-5.1}$ & $13.18^{+0.52}_{-0.69}$ & $8.50^{+0.29}_{-0.26}$ & $8.52^{+0.25}_{-0.22}$ \\
        $K$ (\ms{}) & -- & $130.46^{+3.49}_{-1.59}$ & $53.3^{+5.1}_{-4.7}$ & -- \\
        $e$ & $0.26^{+0.12}_{-0.09}$ & $0.120^{+0.021}_{-0.022}$ & $0.15\pm0.04$ & $0.14^{+0.04}_{-0.05}$ \\
        $\omega$ ($\degree$) & -- & $107.79^{+9.55}_{-3.53}$ & $162\pm8$ & $163^{+9}_{-8}$ \\
        $m\sin i$ ($M_J$) & $40^{+43}_{-28}$ & -- & $5.4^{+0.6}_{-0.5}$ & -- \\
        $i$ ($\degree$) & -- & $57.65^{+8.15}_{-5.75}$ & $66^{+11}_{-9}$ $^{(a)}$ & $71\pm12$ $^{(a)}$ \\
        $\Omega$ ($\degree$) & -- & $62.12^{14.71}_{-12.77}$ & $271^{+15}_{-21}$ $^{(a)}$ & $270^{+30}_{-35}$ $^{(a)}$ \\
        $m$ ($M_J$) & -- & $19.64^{+2.27}_{-2.14}$ & $6.0\pm0.6$ & $5.9^{+0.5}_{-0.4}$ \\
        \hline
        \multicolumn{5}{l}{$^{(a)}$ $i<90\degree$ solution only, for comparison with \citet{Feng2022}.} \\
    \end{tabular}
\end{table*}

\begin{figure*}
    \vspace{5mm}
    \includegraphics[width=\textwidth]{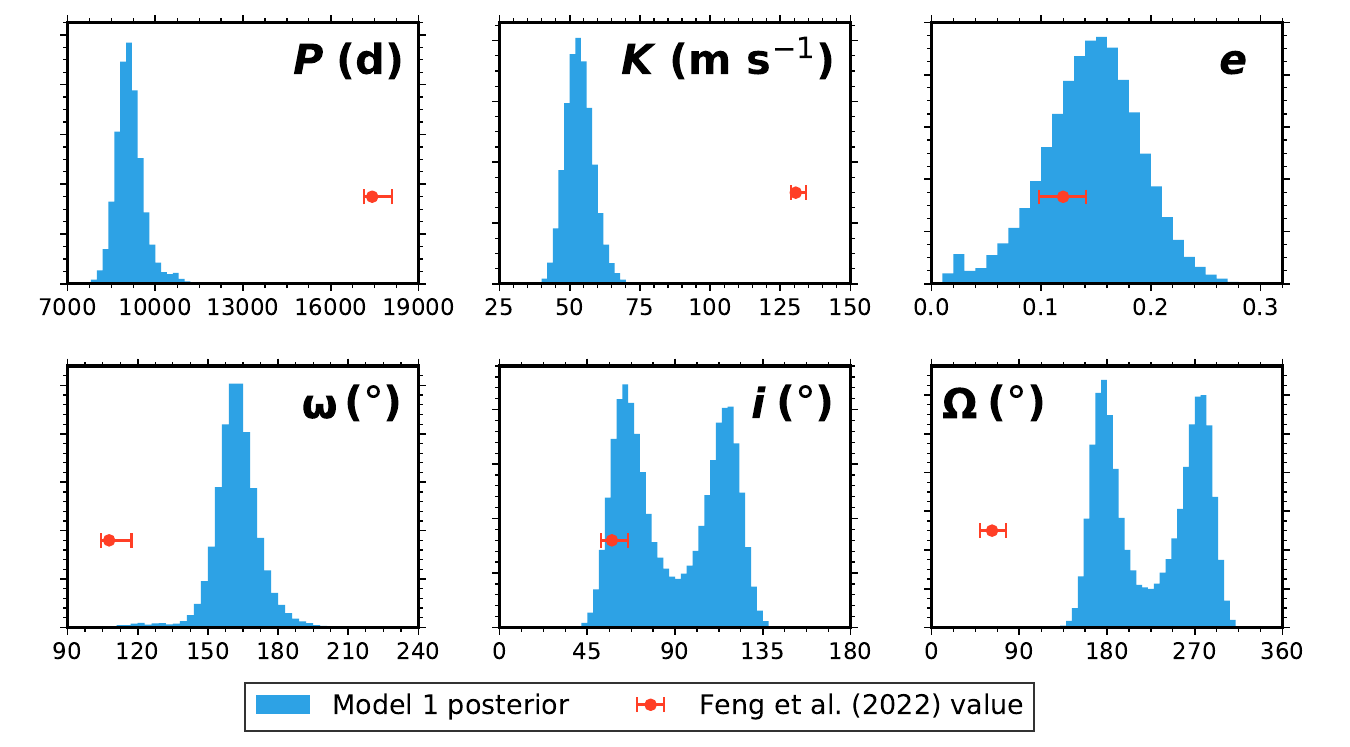}
    \caption{Comparison of our parameter posteriors for \thisstarc{} from Model~1 (histograms) and values from \citet{Feng2022}. Despite similar use of data many of the posteriors differ sharply between these two solutions, and even those parameters which are nominally compatible have confidence intervals of substantially different widths.}
    \label{figure:comparison}
\end{figure*}

\subsubsection{Comparison with previously published results} \label{subsec:why_Feng}

For the outer companion \thisstarc{}, we recover key parameters of $P=9090^{+460}_{-390}$~d, $e=0.15\pm0.04$, and $m=6.0\pm0.6~M_J$ \bmaroon{(a \texttt{corner} plot of all posteriors is shown in Appendix~\ref{appendix:corner})}. We compare our results with those from previous publications in Table~\ref{table:planet_c_comparison}. The parameters initially reported by \citet{Rosenthal2021} for \thisstarc{} are imprecise; $a=15.9^{+7.3}_{-5.1}$~AU, $e=0.26^{+0.12}_{-0.09}$, $m\sin i=40^{+43}_{-28}~M_J$. Our solution in this work stands in reasonable agreement with theirs considering these large uncertainties, with our solution benefiting from the inclusion of additional RVs and astrometry.

However, our parameters differ greatly from those previously published by \citet{Feng2022}, necessitating further scrutiny. In Figure~\ref{figure:comparison} we compare our results to those of \citet{Feng2022} for the key observable parameters of \thisstarc{}. There are large discrepancies for several of the parameters such as $P$ (10$\sigma$) and $K$ (14$\sigma$). Even in cases where our posteriors nominally agree, it is also the uncertainties reported by \citet{Feng2022} are consistently smaller than those found in this work. This contrasts with the good agreement with our alternative fit using \texttt{orvara} (Model~2, Section~\ref{subsec:results_model2}), and we are unable to reproduce the results of \citet{Feng2022} using either of our models.

We first examine whether differences in data can explain these differences between our results. For their fit to the \thisstar{} system, \citet{Feng2022} use RV data from HARPS, HIRES, MIKE, and PFS; in comparison to our model they do not use the published CORALIE and HRS RVs (Section~\ref{subsec:data}). The omission of the CORALIE data may be significant, as only this RV data covers the RV minimum for \thisstarc{}. However this is found to be of secondary importance by our model, as the orbital period of \thisstarc{} is instead mainly constrained by the astrometry.

As in our work, \citet{Feng2022} use data from \textit{Hipparcos} and \textit{Gaia}~(E)DR3 as the astrometric information for their model. However, they fit to the astrometry in a different way to most authors, and it could be argued that this explains of the observed differences. Beginning with the parameterisation in the \textit{Hipparcos} and \textit{Gaia} astrometric solutions in the form of positions and proper motions, \citet{Feng2022} fit directly to both whereas \citet{Brandt2018} and \citet{Kervella2019} reparameterise the data by converting the positional difference between \textit{Hipparcos} and \textit{Gaia} into a time-averaged proper motion, and it is in this basis that most authors prefer to fit the astrometry \citep[e.g.][]{Brandt2019, orvara, Xuan2020, Venner2021, Philipot2023L, Philipot2023S}. \citet{Feng2022} recognise this and assert that their method is superior, claiming that conversion to proper motions is ``biased" without further elaboration \citep[][section~3]{Feng2022}.

To the contrary, however, the orbital information conveyed by the time-averaged proper motion is identical to that which can be gained directly from the positions (i.e. no information is lost in the reparameterisation). This is because a single position measurement provides no information on the orbits of any companions, so it is the difference \textit{over time} between the \textit{Hipparcos} and \textit{Gaia} positions that contains all of the relevant information \citep[on this point see especially][section 3]{Kervella2019}. The assertion that fitting to four position measurements is superior to fitting two time-averaged measurements in proper motion therefore does not withstand scrutiny.\footnote{This is claimed by \citet{Feng2022} in relation to Gliese~229 and 14~Herculis (on which see below). On the contrary, \textit{Hipparcos-Gaia} astrometry is preferable as it includes secondary improvements such as a localised cross-calibration \citep{Brandt2018}.} This difference in parameterisation therefore cannot explain the observed discrepancies.

As our selections of RV and astrometry data are fundamentally similar, we postulate that the discrepancies between our results and those of \citet{Feng2022} are caused by differences between our methods and models. For \thisstar{}, it appears that their model does not reproduce the astrometry well. We have inspected their model figure for this system in \citet[][appendix figure~32]{Feng2022}, and observe that the \textit{Gaia} proper motion lies approximately (+0.35,+0.45)~\masyr{} from the model at the corresponding epoch. Considering the uncertainties listed in Table~\ref{table:astrometry} this represents a $>$10$\sigma$ discrepancy, and significantly contrasts with the good agreement between observations and model that we find in Figure~\ref{figure:astrometry}.

At this point it becomes salient to note that \thisstar{} is not the only system for which there are significant differences between \citet{Feng2022} and other studies. We cite a non-exhaustive selection of examples below:

\begin{itemize}
    \item HD~38529 has two massive planetary companions, of which the outer planet c has been detected with astrometry. \citet{Benedict2010} used \textit{Hubble Space Telescope} (HST) astrometry to measure $i_{\text{c}}=131.7\pm3.7\degree$ and $\Omega_{\text{c}}=218.2\pm7.7\degree$.\footnote{We have applied the angular rotation described by \citet{Xuan2020_disks} to these values. Though the formalism of \citet{Xuan2020} and \citet{Xuan2020_disks} differ in the sign of $\omega$ from most works, for $i$ and $\Omega$ they are consistent \citep{Venner2021}, so the comparison employed here is fair.} Later, \citet{Xuan2020_disks} used \textit{Hipparcos-Gaia} astrometry for the same purpose and found $i_{\text{c}}=135^{+8}_{-14}\degree$ and $\Omega_{\text{c}}=217^{+15}_{-19}\degree$, in very good agreement with \citet{Benedict2010}. However, based on similar data \citet{Feng2022} report $i_{\text{c}}=104.56^{+6.39}_{-8.72}\degree$ and $\Omega_{\text{c}}=37.8^{+16.23}_{-14.91}\degree$, which differ by 3.7$\sigma$ and 11$\sigma$ from the HST result respectively.
    \item 14~Herculis is a system with two long-period giant planets discovered with RVs. \citet{BardalezGagliuffi2021} used \texttt{orvara} to study the orbits of both planets; for the outer planet c, they found parameters including $P_{\text{c}}=144^{+139}_{-58}$~yr, $e_{\text{c}}=0.64^{+0.12}_{-0.13}$, $i_{\text{c}}=101^{+31}_{-33}\degree$, and $m_{\text{c}}=6.9^{+1.7}_{-1.0}~M_J$, while for planet b the relevant parameter here is the orbital inclination $i_{\text{b}}=32.7^{+5.3}_{-3.2}\degree$ which indicates a strong misalignment between the two orbits. Contrasting with these results, \citet{Feng2022} report parameters for 14~Her~c of $P_{\text{c}}=43.07^{+7.27}_{-5.19}$~yr, $e_{\text{c}}=0.393^{+0.045}_{-0.048}$, $i_{\text{c}}=129.10^{+6.26}_{-29.05}\degree$, and $m_{\text{c}}=5.03^{+0.87}_{-1.07}~M_J$, while for planet b they report similar RV parameters but a strongly discordant orbital inclination of $i_{\text{b}}=144.65^{+6.28}_{-3.24}\degree$ (25$\sigma$ from \citealt{BardalezGagliuffi2021}). Most recently \citet{Benedict2023} have performed an independent fit incorporating HST astrometry, finding $P_{\text{c}}=142.8\pm2.8$~yr, $e_{\text{c}}=0.65\pm0.06$, $i_{\text{c}}=82\pm14\degree$, $m_{\text{c}}=7.1^{+1.0}_{-0.6}~M_J$, and $i_{\text{b}}=36\pm3\degree$, which agree only with \citet{BardalezGagliuffi2021}.
    \item The star HD~62364 shows long-term RV variability in HARPS observations. \citet{Feng2022} fit this system with two long-period companions; $P_{\text{b}}=75.02^{+9.78}_{-4.42}$~yr, $e_{\text{b}}=0.863^{+0.012}_{-0.008}$, $m_{\text{b}}=17.44^{+1.62}_{-1.67}~M_J$, and $P_{\text{c}}=204.90^{+22.66}_{-18.79}$~yr, $e_{\text{c}}=0.773\pm0.009$, $m_{\text{c}}=24.93^{+2.68}_{-2.92}~M_J$. This is difficult to understand as the data can be much more simply explained by a single companion with parameters of $P=14.15\pm0.06$~yr, $e=0.6092\pm0.0042$, and $m=17.46^{+0.62}_{-0.59}~M_J$, and there is no need to assume a second companion in the data (\citealt{Xiao2023}; also \citealt{Frensch2023, Philipot2023L}).
    \item Gliese~229~B was one of the first brown dwarfs to be discovered. Based on RVs, imaging, and astrometry from the DR2 HGCA, \citet{Brandt2020} reported the first dynamical mass of $70\pm5~M_J$ for this brown dwarf. Later, with additional data and updated astrometry from the EDR3 HGCA, \citet{BrandtGM2021BD} used \texttt{orvara} to measure an updated and more precise mass of $71.4\pm0.6~M_J$ for Gliese~229~B. However, based on largely the same data, \citet{Feng2022} report $m=60.4^{+2.3}_{-2.4}~M_J$ (4.6$\sigma$ lower). It is not possible to reconcile this mass with \texttt{orvara} results (G.~M. Brandt, private communication).
    \item HD~211847 was found to have a candidate brown dwarf companion from CORALIE RVs by \citet{Sahlmann2011}. However, this companion was subsequently resolved in SPHERE imaging by \citet{Moutou2017}, identifying HD~211847~B as an M-dwarf observed at a near-polar orbital inclination. Seemingly unaware of this result, \citet{Feng2022} claim that HD~211847~B is a brown dwarf at $a=4.51^{+0.29}_{-0.46}$~AU with $m=55.32^{+1.34}_{-18.49}~M_J$. This disagrees sharply with \citet{Xiao2023} and \citet{Philipot2023S}, who incorporate the imaging detection into their fits and find $a=\big(6.83\pm0.06,\,6.78\pm0.08\big)$~AU and $m=\big(148.6^{+3.7}_{-3.6},\,148\pm5\big)~M_J$ respectively. The latter values for the mass agree with the photometric estimate of $155\pm9~M_J$ \citep{Moutou2017}. The mass estimate of \citet{Feng2022} is therefore discrepant by 11$\sigma$, 24$\sigma$, and 18$\sigma$ from the values of \citet{Moutou2017}, \citet{Xiao2023}, and \citet{Philipot2023S}, respectively.
    \item Gliese~680 and HD~111031 are two stars with RV accelerations for which \citet{Feng2022} claim brown dwarf companions with $a=\big(10.14^{+1.84}_{-1.70},\,13.10^{+0.75}_{-1.09}\big)$~AU and $m=\big(25.10^{+6.16}_{-11.15},\,54.17^{+5.32}_{-6.15}\big)~M_J$ respectively. However, in both cases these companions have previously been resolved as M-dwarfs in imaging observations \citep{WardDuong2015, Gonzales2020, Dalba2021}. Incorporating the imaging data into joint fits, \citet{Philipot2023L} find very different parameters of $a=\big(32^{+9}_{-6},\,21.1\pm0.6\big)$~AU and $m=\big(186\pm4,\,135\pm3\big)~M_J$ respectively. \citet{Feng2022} therefore underestimate the masses of these companions by 21$\sigma$ and 13$\sigma$ respectively.
\end{itemize}

The above list is not exhaustive, but is sufficient to give an impression of the frequency and scope of disagreements between \citet{Feng2022} and other works. These go beyond any conceivable statistical explanation, as evidenced by the number of $>$10$\sigma$ discrepancies listed above. These disagreements extend to many different works using many different fitting methods, chiefly based on \textit{Hipparcos-Gaia} astrometry but also including HST astrometry. It is also significant that for the systems where there are multiple independent sources of comparison such as 14~Her or HD~211847, the alternate solutions are mutually consistent with each other yet disagree with \citet{Feng2022}; this is repeated by our two consistent sets of results for \thisstar{}. This demonstrates that the models causing this tension are those of \citet{Feng2022}, rather than those of other authors.

It is not the case that \textit{all} solutions from \citet{Feng2022} are inconsistent with results from other works; consider \citet[][section 6.2]{Xiao2023}. However, the solutions showing the strongest discrepancies tend to be those longest orbital periods (typically $P>10$~yr), which is also the case for \thisstarc{}. This leads us to speculate that there is a period dependence in the origin of discrepant solutions in \citet{Feng2022}. \bmaroon{We intend to elaborate on this elsewhere (Venner~et~al.~in~prep).}

To conclude, we find that our solution for \thisstarc{} differs significantly from that of \citet{Feng2022}, and that this is paralleled by similar cases for other planets, brown dwarfs, and stars from the literature. Discrepancies chiefly occur for companions with comparatively long orbital periods, and the circumstances of these difficulties are sufficient to demonstrate that it is the solutions from \citet{Feng2022} which are in exception to other works rather than vice-versa. This indicates that there is some issue with the method of \citet{Feng2022} which has resulted in inaccurate orbital solutions.\footnote{This may explain the observation of \citet{Benedict2023} that the inclination distribution from \citet{Feng2022} does not match the distribution expected for random viewing orientations.}

\begin{figure*}
	\includegraphics[width=\textwidth]{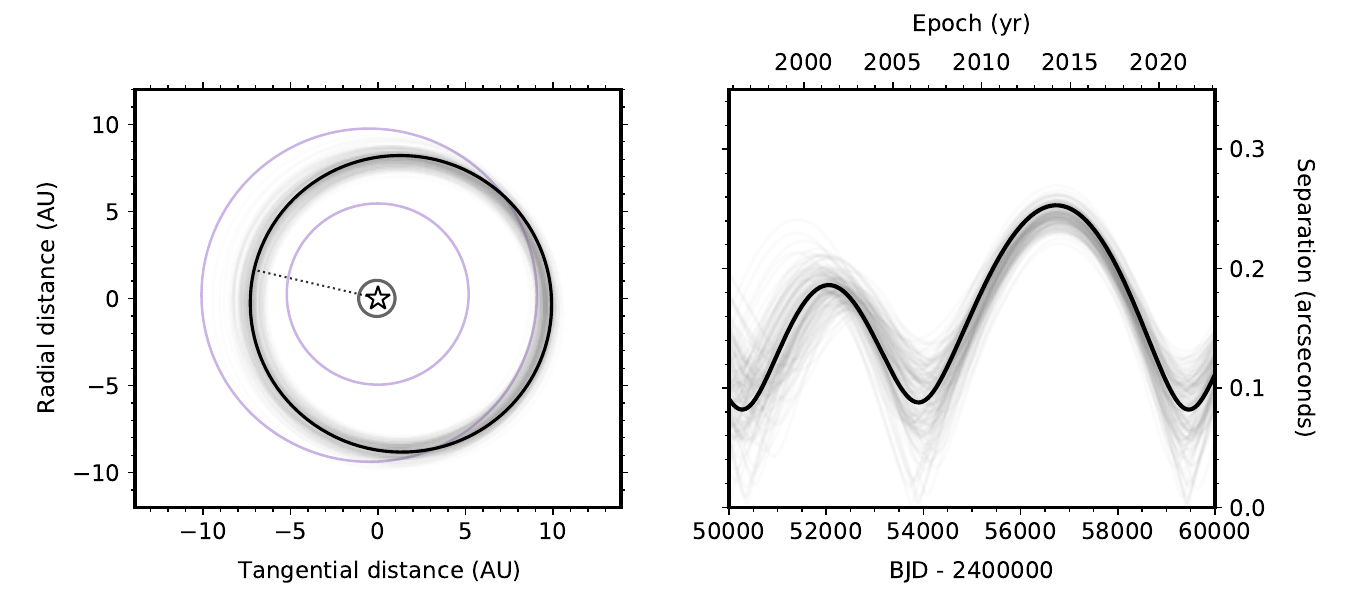}
	\caption{(\textit{Left}) Plan view of the orbit of \thisstarc{}. For comparison we plot the orbits of Jupiter and Saturn (\bmaroon{purple}). The orbit of \thisstarb{} \bmaroon{is} also shown \bmaroon{to scale}
    ; the planetary orbits are assumed to be coplanar. With $a=8.50^{+0.29}_{-0.26}$~AU and $e=0.15\pm0.04$, \thisstarc{} has an orbit reminiscent of Saturn. (\textit{Right}) Planet-star separation over the observed orbit of \thisstarc{}. Though its semi-major axis is exceptionally large among indirectly detected exoplanets, the age and distance of \thisstarc{} means that it is too faint and too close to its star to be detected with modern imaging instruments. However, it is a potential target for imaging with future 30m telescopes.}
	\label{figure:planet_c}
\end{figure*}

\subsubsection{HD 28185 c as a massive Saturn analogue} \label{subsec:planet_c_discussion}

\thisstarc{} is a significant addition to the sample of long-period exoplanets. At $P=9090^{+460}_{-390}$~d \big($24.9^{+1.3}_{-1.1}$~yr\big), \thisstarc{} has among the longest orbital period of any exoplanet for which this parameter has been precisely constrained; furthermore, with $8.50^{+0.29}_{-0.26}$~AU and $e=0.15\pm0.04$, the orbit of this planet invites comparisons to Saturn ($P=29.45$~yr; $a=9.58$~AU; $e=0.057$). However, its mass of $6.0\pm0.6~M_J$ is larger than that of Saturn by a factor of 20. Like the inner planet~b, \thisstarc{} is therefore a ``Super-Jupiter."

In Figure~\ref{figure:planet_c} we visualise the orbit of \thisstarc{} with comparison to the orbits of Jupiter and Saturn in the solar system, along with the interior orbit of \thisstarb{}. Though the orbit of \thisstarc{} is somewhat smaller in scale than that of Saturn at apoastron their separations are approximately equal, and the exoplanet never reaches interior to the orbit of Jupiter ($a=5.20$~AU). \thisstarc{} is therefore very much a massive analogue of Saturn.

We next place \thisstarc{} in the context of the known long-period giant planets. We assemble the sample of exoplanets with well-constrained orbits with periods beyond $P>5500$~d ($\approx$15~yr), with this cutoff corresponding approximately to half the orbital period of Saturn. This effectively circumscribes the known exoplanets with periods substantially longer than that of Jupiter ($P=11.86$~yr). A total of 37 exoplanets pass our selection, and we summarise their properties in Appendix~\ref{appendix:planet_table}. Despite the difficulty inherent to the detection of these long-period planets a majority were originally detected with RVs; in turn a majority of these planets have been further characterised using \textit{Hipparcos-Gaia} astrometry, testifying to the significant impact this data has had on the study of long-period giant planets in five years since its inception \citep{Brandt2018, Kervella2019}. In contrast our requirement of well-constrained orbits excludes most of the exoplanets current known from direct imaging, leaving only those with relatively close orbits ($a<30$~AU) in this sample. Nonetheless it is notable that the period ranges of the directly- and indirectly-detected exoplanets increasingly overlap, and it is particularly remarkable how the parameters of \thisstarc{} bear resemblance to the famous direct imaging planets $\beta$~Pictoris~b \big(\citealt{Lagrange2009, Lagrange2010}; $P=8864^{+118}_{-113}$~d, $e=0.119\pm0.008$, $m=9.3^{+2.6}_{-2.5}~M_J$, \citealt{BrandtGM2021}\big) and 51~Eridani~b \big(\citealt{Macintosh2015}; $P=9100^{+1100}_{-1500}$~d, $e=0.57^{+0.08}_{-0.06}$, $m=3.1^{+0.5}_{-0.7}~M_J$, \citealt{Dupuy2022}\big). The outer detection limits of RVs and the inner limits of direct imaging currently lie at $\sim$10~AU for the typical targets of these methods \citep{Fulton2021, Nielsen2019, Vigan2021}, so this highlights how these methods are increasingly sampling a complementary sample of planets at Saturn-like orbital separations.

The similarity between \thisstarc{} and these famous direct imaging planets motivates consideration of this planet's viability for imaging. Unfortunately this \bmaroon{is significantly impeded by} the great age of the system ($8.3\pm1.0$~Gyr, Table~\ref{table:parameters}) \bmaroon{and the modest projected separation, which never extends further than $\sim$0.25" (Figure~\ref{figure:planet_c}). Using the \texttt{ATMO 2020} atmosphere models \citep{Phillips2020}, we estimate that \thisstarc{} has an effective temperature of $\sim$230~K and a $H$-band planet-star contrast of $\approx$10$^{-8}$, which is considerably beyond the detection capabilities of current near-infrared imaging instruments. The issue of contrast decreases towards the mid-infrared as the contribution of planetary thermal emission increases, such that the contrast reaches $\approx$10$^{-4}$ at 10~$\mu$m. However, the planet cannot plausibly be detected with JWST/MIRI as it is always interior to the coronagraph inner working angle \citep[$0.3-0.5"$ depending on the filter;][]{Boccaletti2015}.}

\bmaroon{The prospects for direct detection improve significantly once 30m-class telescope instruments are brought into consideration. Using ELT/METIS as an example, $\sim$10$^{-8}$ contrast in $L$-band (3.5~$\mu$m) and $\sim$10$^{-6}$ contrast in $M$-band (4.8~$\mu$m) are intended \citep{Brandl2014}. We estimate the corresponding contrasts for our target to be $\approx$10$^{-7}$ and $\approx$10$^{-4}$ respectively. Thus, while \thisstarc{} lies beyond the capabilities of current technology, it may be a suitable target for direct imaging with future 30m-class imaging instruments.}

\begin{figure*}
	\includegraphics[width=\textwidth]{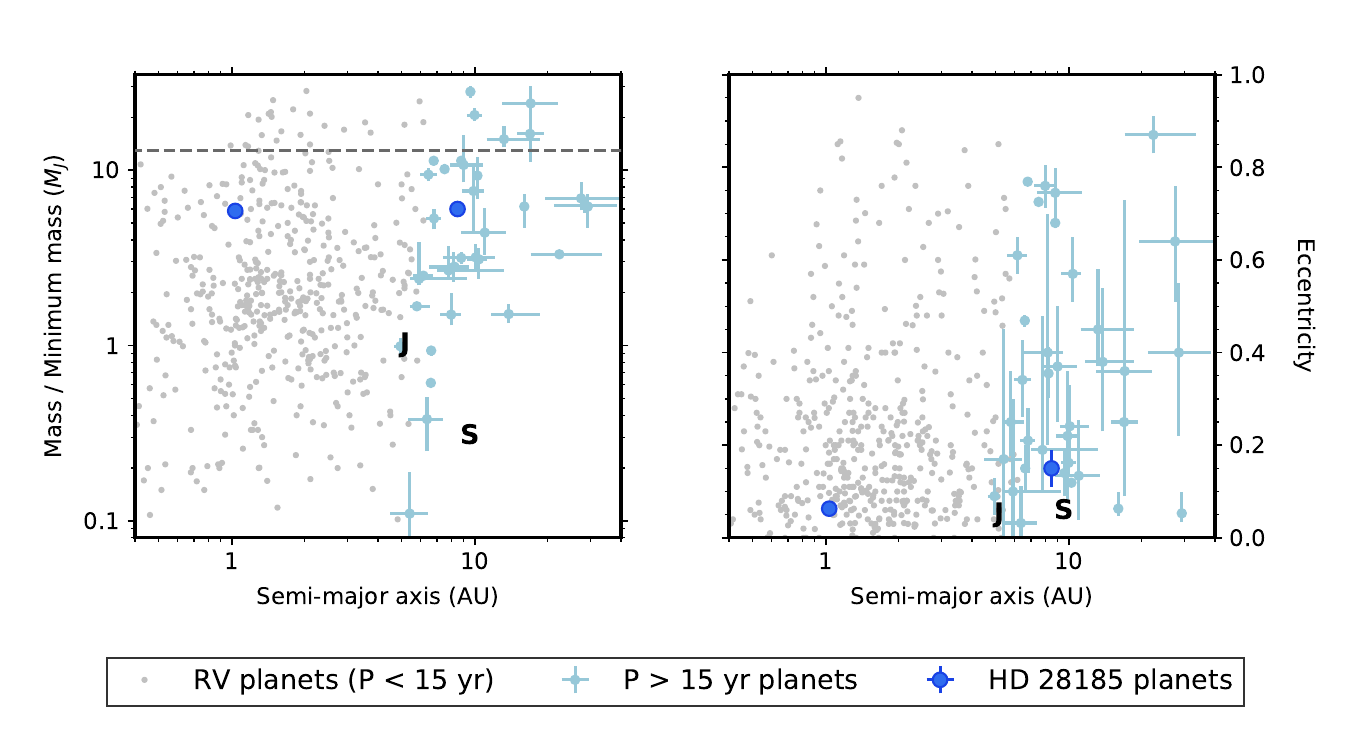}
	\caption{The \thisstar{} planets placed in the context of known planets. Known long-period planets with $P>15$~yr are highlighted, whereas shorter-period planets from RV surveys are shown in grey (without error bars). We also label the locations of Jupiter and Saturn. (\textit{Left}) Planet mass (or minimum mass) against semi-major axis; the location of the deuterium-burning limit, the traditional delimitation between planets and brown dwarfs, at $13~M_J$ is marked by the dotted line. (\textit{Right}) Orbital eccentricity against semi-major axis. While \thisstarb{} lies in a well-represented area of parameter space, \thisstarc{} has only a small number of peers. It also has among the best-constrained parameters of any exoplanet in this region.}
	\label{figure:population}
\end{figure*}

In Figure~\ref{figure:population} we elaborate on the context of the \thisstar{} planets by plotting their mass and eccentricity against semi-major axis along with the long-period exoplanets selected in Appendix~\ref{appendix:planet_table}, other shorter-period exoplanets detected with RVs for specific comparison with \thisstarb{}, and Jupiter and Saturn from the solar system. As remarked in Section~\ref{subsec:planet_b}, \thisstarb{} lies in a region of parameter space richly populated by jovian planets with $\approx$1~AU orbits. In contrast there is a comparatively small number of peers for \thisstarc{}, undoubtedly reflecting the difficulty in the detection of exoplanets on such distant orbits. Though the total mass of the \thisstar{} planetary system is relatively large ($\Sigma(m)\geq12~M_J$), \thisstarb{}~and~c both retain low, solar system-like orbital eccentricities; \thisstarc{} in particular has one of the lowest eccentricities of any known exoplanet within its semi-major axis range, and contrasts with the substantial number of planets in the same region with $e\gtrapprox0.3$. This is indicative of a dynamically quiescent formation history for the \thisstar{} system.

\thisstarc{} is an important addition to the sample of long-period giant planets from RV surveys. The frequency of giant planets at such wide orbital separations is not yet well-constrained; the addition of \thisstarc{} to this group will help to constrain the occurrence rate of giant planets beyond the ice line, for which the occurrence rate function against orbital separation is not yet settled \citep{Fernandes2019, Wittenmyer2020, Fulton2021, Lagrange2023}. The discovery of other planets on similarly distant orbits with RVs, astrometry, and direct imaging will eventually lead to a complete picture of giant planet occurrence beyond the ice line.

\section{Conclusions} \label{sec:conclusions}

\thisstar{} is a Sun-like star found to have a giant planet on an Earth-like orbit by \citet{Santos2001}, and suspected to have a more distant second companion since at least \citet{Chauvin2006}. Recently this outer companion has been claimed as a brown dwarf \citep{Rosenthal2021, Feng2022}. In this work we have revisited the companions of \thisstar{} using published RV observations over a duration of 22~yr and \textit{Hipparcos-Gaia} astrometry spanning 25~yr. We confirm the known properties of \thisstarb{} with high precision, finding key parameters of $P_{\text{b}}=385.92^{+0.06}_{-0.07}$~d, $e_{\text{b}}=0.063\pm0.004$, and $m\sin i_{\text{b}}=5.85\pm0.08~M_J$, and identify this planet as the archetype of the now-numerous population of temperature jovians. For the outer companion \thisstarc{} we find tightly constrained parameters of $P=9090^{+460}_{-390}$~d \big($24.9^{+1.3}_{-1.1}$~yr\big), $e=0.15\pm0.04$, and $m=6.0\pm0.6~M_J$. We recover consistent results for this companion from two independent models, yet these values differ substantially from those found by \citet{Feng2022}. In particular, our results are the first to return a planetary mass for \thisstarc{}. We establish that this companion is a massive analogue of Saturn, having one of the longest orbital periods of any exoplanet found with indirect detection techniques. We highlight strong discrepancies between orbital solutions from \citet{Feng2022} and those of other works.

\section*{Acknowledgements}

We acknowledge and pay respect to Australia’s Aboriginal and Torres Strait Islander peoples, who are the traditional custodians of the lands, waterways and skies all across Australia. We thank the anonymous referee for providing many comments that have helped to improve this work. AV is supported by ARC DECRA project DE200101840. Based on observations made with ESO Telescopes at the La Silla Paranal Observatory under programme ID 072.C-0488. This research has made use of the SIMBAD database and VizieR catalogue access tool, operated at CDS, Strasbourg, France. This research has made use of NASA's Astrophysics Data System. This research has made use of the Washington Double Star Catalog maintained at the U.S. Naval Observatory. This research has made use of the NASA Exoplanet Archive, which is operated by the California Institute of Technology, under contract with the National Aeronautics and Space Administration under the Exoplanet Exploration Program. This work has made use of data from the European Space Agency (ESA) mission {\it Gaia} (\url{https://www.cosmos.esa.int/gaia}), processed by the {\it Gaia} Data Processing and Analysis Consortium (DPAC, \url{https://www.cosmos.esa.int/web/gaia/dpac/consortium}). Funding for the DPAC has been provided by national institutions, in particular the institutions participating in the {\it Gaia} Multilateral Agreement.

\section*{Data Availability}

All data used in this work has been collated from publicly accessible
depositories. The sources of the radial velocities and astrometry, which form the bulk of the data used in our analysis, are described in Section~\ref{subsec:data}.



\bibliographystyle{mnras}
\bibliography{bib} 




\appendix

\section[orvara figures]{\texttt{orvara} figures} \label{appendix:orvara}

\begin{figure*}
	\includegraphics[width=0.35\textheight]{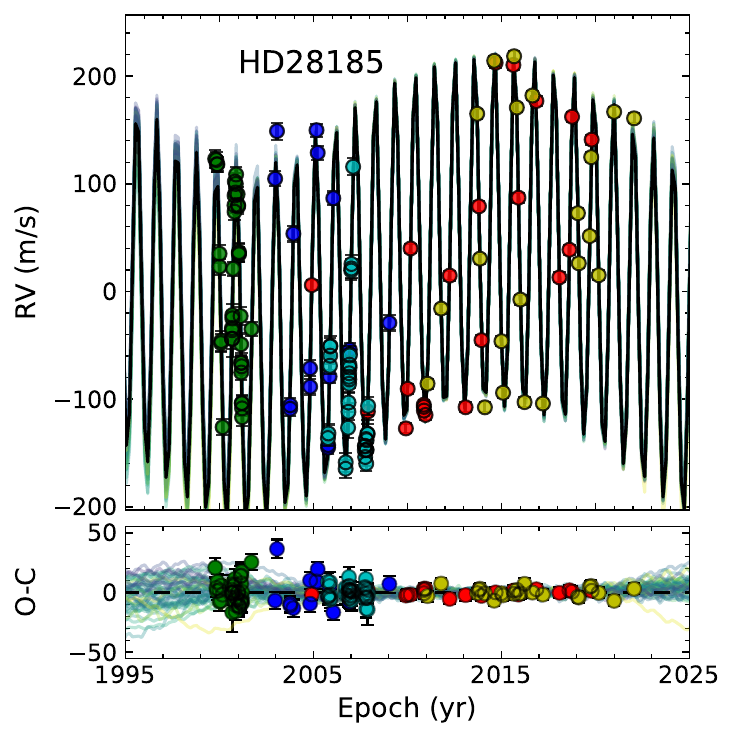}
	\includegraphics[width=0.35\textheight]{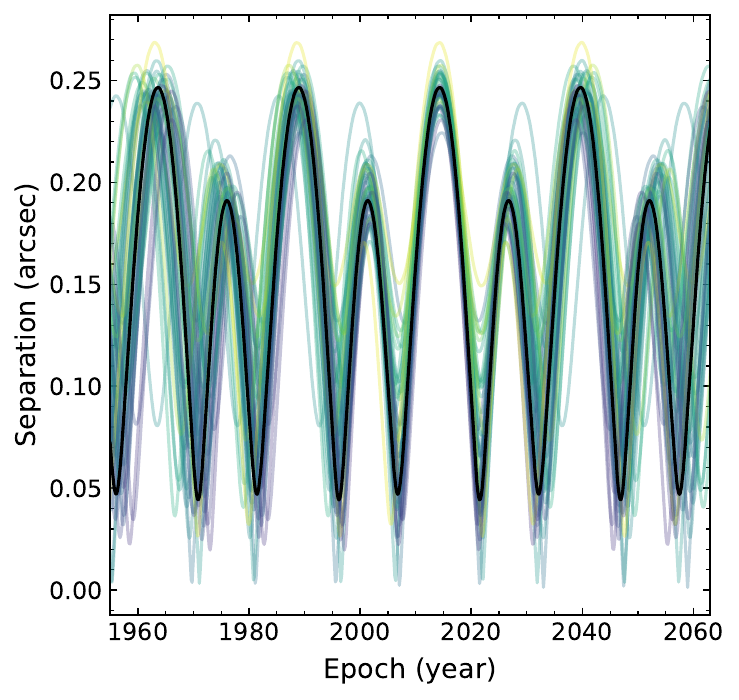}
	\includegraphics[width=0.35\textheight]{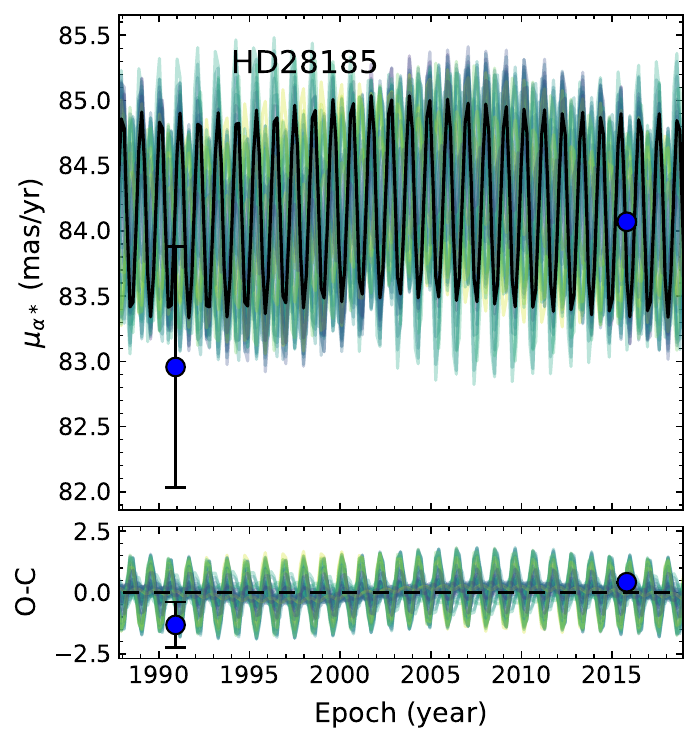}
	\includegraphics[width=0.35\textheight]{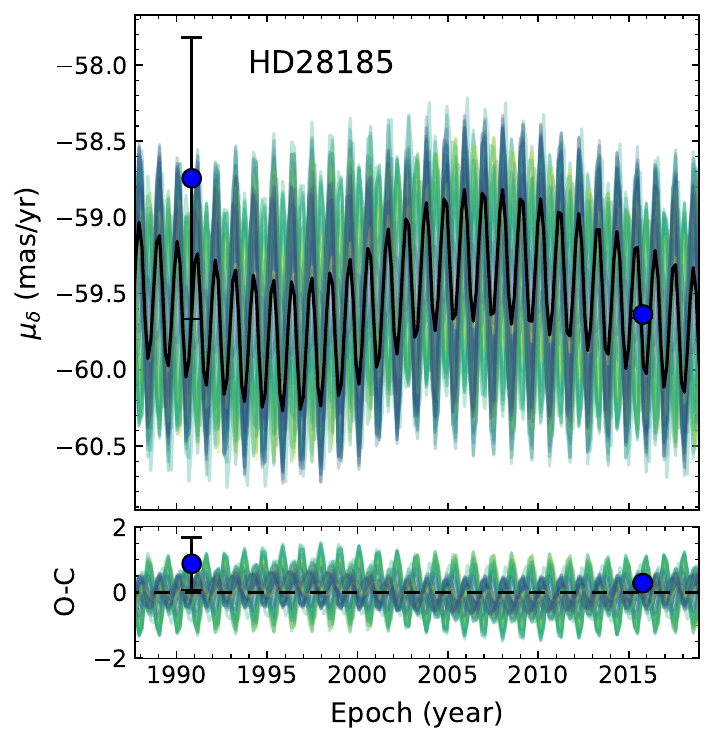}
	\caption{Figures from our \texttt{orvara} model (Model 2; see Section~\ref{subsec:results_model2}). (\textit{upper left}) Radial velocities; compare Figure~\ref{figure:RV}. (\textit{upper right}) Projected separation; compare Figure~\ref{figure:planet_c}. (\textit{lower left, lower right}) \textit{Hipparcos-Gaia} astrometry; compare Figure~\ref{figure:astrometry}, and observe that this differs in fitting for the astrometric signal of \thisstarb{}.}
	\label{figure:orvara}
\end{figure*}

In Figure~\ref{figure:orvara} we show plots from our \texttt{orvara} solution (Model~2, Section~\ref{subsec:results_model2}).

\section{Corner figure} \label{appendix:corner}

\begin{figure*}
    \centering
    \includegraphics[width=\textwidth]{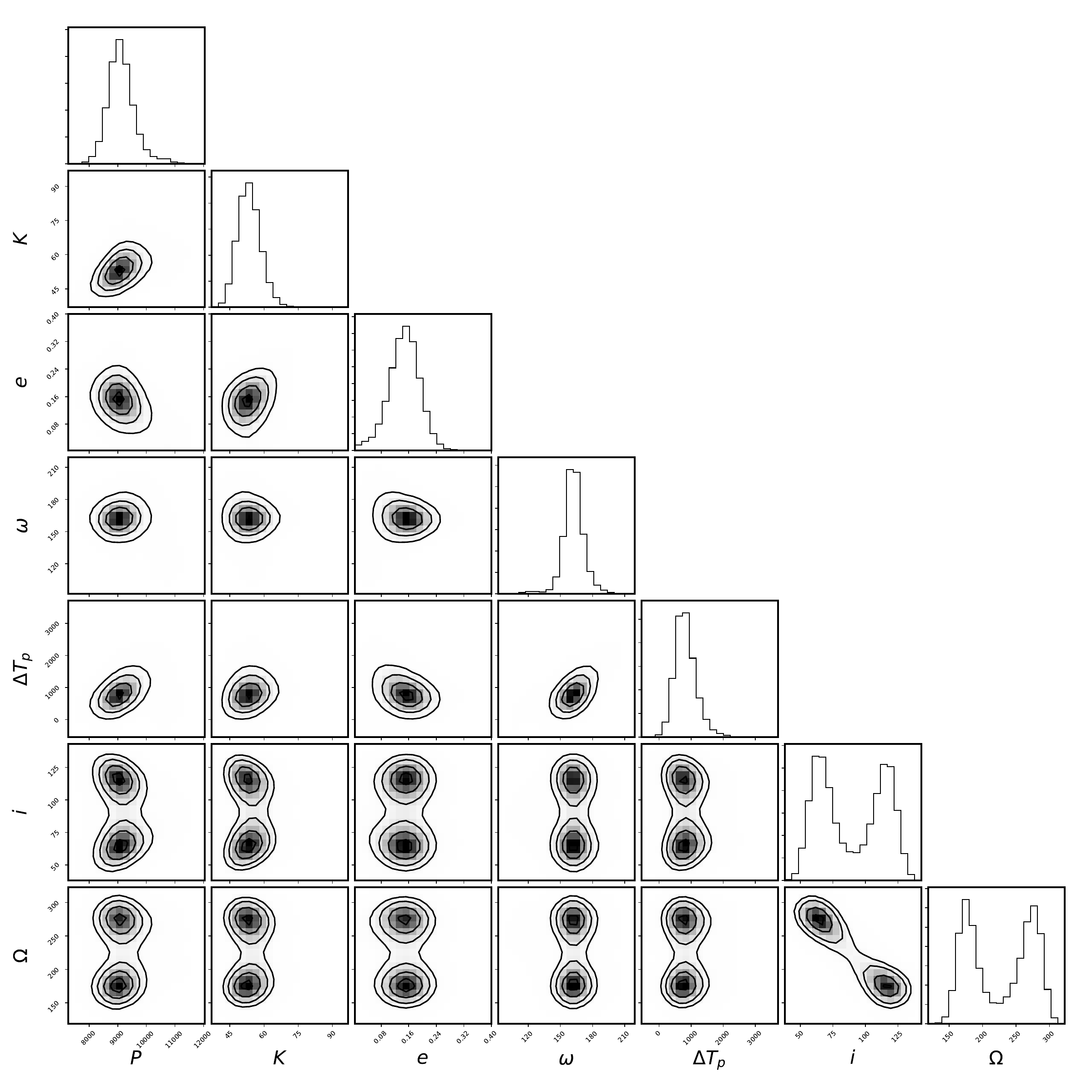}
    \caption{\texttt{corner} plot \citep{corner} of the posteriors for \thisstarc{}.}
    \label{figure:corner}
\end{figure*}

In Figure~\ref{figure:corner} we provide a \texttt{corner} plot for the modelled parameters of \thisstarc{} from Model~1 (Section~\ref{subsec:results_model1}). This figure is a superset of Figure~\ref{figure:bimodal}, which includes only the orbital inclination $i$ and longitude of node $\Omega$.

\section{Comparative long-period planet sample} \label{appendix:planet_table}

\begin{table*}
\centering
\caption{Long-period ($P>5500$~d) exoplanet sample. \thisstarc{} is highlighted in bold.}
\label{table:long_period_planets}
\begin{tabular}{lllllcll}
\hline
Name & Period (d) & $a$ (AU) & Eccentricity & Mass ($M_J$) & $m\sin i$? & Method & Reference \\
\hline
HD 34445 g & $5700\pm1500$ & $6.36\pm1.02$ & $0.032^{+0.08}_{-0.032}$ & $0.38\pm0.13$ & Y & RV & \citet{Vogt2017} \\
HD 133131 B b & $5769\pm831$ & $6.15\pm0.59$ & $0.61\pm0.04$ & $2.50\pm0.05$ & Y & RV & \citet{Teske2016} \\
HD 134987 c & $5960^{+170}_{-150}$ & $6.62^{+0.16}_{-0.15}$ & $0.15\pm0.05$ & $0.935\pm0.06$ & Y & RV & \citet{Rosenthal2021} \\
Gliese 849 c & $5990^{+110}_{-100}$ & $4.95^{+0.25}_{-0.28}$ & $0.09\pm0.04$ & $0.99\pm0.11$ & Y & RV & \citet{Pinamonti2023} \\
HD 142 c & $6005\pm447$ & $6.8\pm0.5$ & $0.21\pm0.07$ & $5.3\pm0.7$ & Y & RV & \citet{Wittenmyer2012} \\
HD 219077 b & $6199^{+52}_{-46}$ & $6.78\pm0.08$ & $0.769\pm0.002$ & $11.3\pm0.4$ &  & RV + HG & \citet{Philipot2023S} $^{(a)}$ \\
HIP 10337 c & $6360^{+6260}_{-711}$ & $5.9^{+3.4}_{-0.5}$ & $0.1^{+0.2}_{-0.1}$ & $2.4^{+1.5}_{-0.2}$ & Y & RV & \citet{Frensch2023} \\
HD 25015 b & $6360^{+770}_{-690}$ & $6.45^{+0.52}_{-0.47}$ & $0.341^{+0.086}_{-0.080}$ & $9.42^{+0.85}_{-0.78}$ &  & RV + HG & \citet{Xiao2023} \\
HD 181433 d & $7012\pm276$ & $6.60\pm0.22$ & $0.469\pm0.013$ & $0.612\pm0.004$ & Y & RV & \citet{Horner2019} \\
HD 4203 c & $7400^{+8900}_{-1100}$ & $7.8^{+5.4}_{-0.8}$ & $0.19^{+0.29}_{-0.09}$ & $2.68^{+0.99}_{-0.24}$ & Y & RV & \citet{Rosenthal2021} \\
HD 181234 b & $7450^{+120}_{-100}$ & $7.52^{+0.16}_{-0.17}$ & $0.7254\pm0.0094$ & $10.13^{+0.74}_{-0.63}$ &  & RV + HG & \citet{Xiao2023} \\
Gliese 317 c & $7500^{+1500}_{-720}$ & $5.78^{+0.75}_{-0.38}$ & $0.25^{+0.11}_{-0.074}$ & $1.673^{+0.078}_{-0.076}$ & Y & RV & \citet{Rosenthal2021} \\
Gliese 15 A c & $7600^{+2200}_{-1700}$ & $5.4^{+1.0}_{-0.9}$ & $0.27^{+0.28}_{-0.19}$ & $0.11^{+0.08}_{-0.06}$ & Y & RV & \citet{Pinamonti2018} \\
AF Lep b & $8150^{+2050}_{-2450}$ & $8.2^{+1.3}_{-1.7}$ & $0.4^{+0.3}_{-0.2}$ & $2.8^{+0.6}_{-0.5}$ &  & Imaging + HG & \citet{Zhang2023} \\
HD 83443 c & $8241^{+1019}_{-530}$ & $8.0\pm0.8$ & $0.760^{+0.046}_{-0.047}$ & $1.5^{+0.5}_{-0.2}$ &  & RV + HG & \citet{Errico2022} \\
HD 92788 c & $8360^{+540}_{-390}$ & $8.26^{+0.36}_{-0.28}$ & $0.355^{+0.057}_{-0.052}$ & $2.81^{+0.18}_{-0.17}$ & Y & RV & \citet{Rosenthal2021} \\
HAT-P-2 c & $8500^{+2600}_{-1500}$ & $9.0^{+1.8}_{-1.1}$ & $0.37^{+0.13}_{-0.12}$ & $10.7^{+5.2}_{-2.2}$ &  & RV + HG & \citet{deBeurs2023} \\
$\beta$ Pic b & $8864^{+118}_{-113}$ & $10.26\pm0.10$ & $0.119\pm0.008$ & $9.3^{+2.6}_{-2.5}$ &  & Imaging + RV + HG & \citet{BrandtGM2021} \\
\textbf{HD 28185 c} & $9090^{+460}_{-390}$ & $8.50^{+0.29}_{-0.26}$ & $0.15\pm0.04$ & $6.0\pm0.6$ &  & RV + HG & This work \\
51 Eri b & $9100^{+1100}_{-1500}$ & $10.4^{+0.8}_{-1.1}$ & $0.57^{+0.08}_{-0.06}$ & $3.1^{+0.5}_{-0.7}$ &  & Imaging + HG & \citet{Dupuy2022} \\
HD 206893 b & $9350\pm440$ & $9.6^{+0.4}_{-0.3}$ & $0.14\pm0.05$ & $28.0^{+2.2}_{-2.1}$ &  & Imaging + RV + HG & \citet{Hinkley2022} \\
HD 190984 b & $9970^{+4380}_{-2200}$ & $8.8^{+2.5}_{-1.4}$ & $0.745^{+0.054}_{-0.047}$ & $3.16^{+0.25}_{-0.26}$ &  & RV + HG & \citet{Xiao2023} \\
HD 221420 b & $10120^{+1100}_{-910}$ & $9.99^{+0.74}_{-0.70}$ & $0.162^{+0.035}_{-0.030}$ & $20.6^{+2.0}_{-1.6}$ &  & RV + HG & \citet{Li2021} \\
HD 16905 b  & $10256^{+618}_{-522}$ & $8.8^{+0.4}_{-0.3}$ & $0.68^{+0.02}_{-0.01}$ & $11.3^{+0.6}_{-0.7}$ &  & RV + HG & \citet{Philipot2023L} \\
HD 50499 c & $10400^{+3200}_{-1300}$ & $10.1^{+2.0}_{-0.9}$ & $0.241^{+0.089}_{-0.075}$ & $3.18^{+0.63}_{-0.46}$ & Y & RV & \citet{Rosenthal2021} \\
HD 30177 c & $11613\pm1837$ & $9.89\pm1.04$ & $0.22\pm0.14$ & $7.6\pm3.1$ & Y & RV & \citet{Wittenmyer2017} \\
HD 73267 c & $13900^{+5100}_{-4000}$ & $11.0^{+2.5}_{-2.2}$ & $0.134^{+0.120}_{-0.095}$ & $4.4^{+1.7}_{-1.1}$ &  & RV + HG & \citet{Xiao2023} \\
HD 68988 c & $16000^{-11000}_{-3500}$ & $13.2^{+5.3}_{-2.0}$ & $0.45^{+0.13}_{-0.08}$ & $15.0^{+2.8}_{-1.5}$ & Y & RV & \citet{Rosenthal2021} \\
29 Cyg b & $18600^{+6200}_{-2800}$ & $16.9^{+2.4}_{-1.9}$ & $0.25^{+0.14}_{-0.16}$ & $16.1^{+5.4}_{-5.0}$ &  & Imaging + HG & \citet{Currie2023} \\
47 UMa d & $19000^{+11000}_{-4000}$ & $13.8^{+4.8}_{-2.1}$ & $0.38^{+0.16}_{-0.15}$ & $1.51^{+0.22}_{-0.17}$ & Y & RV & \citet{Rosenthal2021} \\
HR 8799 e & - & $16.0^{+0.5}_{-0.6}$ & $0.063^{+0.037}_{-0.015}$ & $6.2^{+1.1}_{-1.5}$ $^{(b)}$ &  & Imaging & \citet{Sepulveda2022} \\
HD 28736 b & $22000^{+11000}_{-5800}$ & $17^{+5}_{-4}$ & $0.36^{+0.37}_{-0.25}$ & $24^{+6}_{-4}$ &  & Imaging + RV + HG & \citet{Franson2022} \\
HD 111232 c & $32100^{+13500}_{-9900}$ & $18.8^{+5.0}_{-4.1}$ & $0.33^{+0.10}_{-0.09}$ & $20.7^{+3.4}_{-3.2}$ &  & RV + HG & \citet{Xiao2023} \\
HR 5183 b & $37200^{+30700}_{-12400}$ & $22.3^{+11.0}_{-5.3}$ & $0.87\pm0.04$ & $3.31^{+0.18}_{-0.14}$ &  & RV + HG & \citet{Venner2022} \\
14 Her c & $53000^{+51000}_{-21000}$ & $27.4^{+16}_{-7.9}$ & $0.64^{+0.12}_{-0.13}$ & $6.9^{+1.7}_{-1.0}$ &  & RV + HG & \citet{BardalezGagliuffi2021} \\
$\epsilon$ Ind b & - & $28.4^{+10}_{-7.2}$ & $0.40^{+0.15}_{-0.18}$ & $6.3\pm0.6$ &  & Imaging + RV + HG & \citet{Matthews2024} \\
HR 8799 d & - & $29.2^{+1.3}_{-1.4}$ & $0.053^{+0.047}_{-0.019}$ & $6.2^{+1.1}_{-1.5}$ $^{(b)}$ &  & Imaging & \citet{Sepulveda2022} \\
\hline
\multicolumn{8}{l}{$^{(a)}$ In \citet[][table 3]{Philipot2023S}, $P$, $a$, and $e$ for HD~219077~b appear to be switched with those of HD~211847~B. We use the values as attributed in the text.} \\
\multicolumn{8}{l}{$^{(b)}$ $\text{[Fe/H]}=0.0$, normal mass prior solution from \citet{Sepulveda2022}.} \\
\end{tabular}
\end{table*}

In Table~\ref{table:long_period_planets} we present the sample of long-period planets with well-constrained orbits used in Figure~\ref{figure:population}. To assemble this list we began by querying the NASA Exoplanet Archive,\footnote{\url{https://exoplanetarchive.ipac.caltech.edu/}, accessed 2023-08-20.} and then supplemented this with relevant literature results not present in the archive. For our selection we restricted our sample to planets with $P>5500$~d ($\approx$15~yr, or approximately half the orbital period of Saturn). We then manually excluded the planets with imprecise orbits identified as problematic by \citet{Lagrange2023}, e.g. HD~26161~b ($a=20.4^{+7.9}_{-4.9}$~AU, $m=13.5^{+8.5}_{-3.7}~M_J$, \citealt{Rosenthal2021}; see \citealt{Lagrange2023} for reservations on this target and other imprecise orbits). \bmaroon{We also exclude certain spurious planets (Venner et al. in prep).} We also enforce an upper semi-major axis limit of $a<30$~AU for Figure~\ref{figure:population}, though this makes little difference to the sample selection as there are very few planets with well-constrained orbits beyond this.

As mentioned in Section~\ref{subsec:planet_c_discussion}, 37 exoplanets pass this selection (including \thisstarc{}). In Table~\ref{table:long_period_planets} we list their names, orbital periods, semi-major axes, eccentricities, and masses, with a flag indicating whether the latter is a minimum mass ($m\sin i$). We then provide the reference for the quoted parameters and, in this case solely for the interest of the reader, a summary of the types of data used for the orbital solution in the ``Method" column.


\bsp	
\label{lastpage}
\end{document}